\documentclass[superscriptaddress, twocolumn]{revtex4-1}

\usepackage{graphicx,color}
\usepackage{textcomp}
\usepackage[squaren, Gray, cdot]{SIunits}
\usepackage{amsmath}
\DeclareMathOperator{\Tr}{Tr}
\usepackage{natbib}
\usepackage{revsymb}
\usepackage[colorlinks=true,linkcolor=blue,citecolor=blue,urlcolor=blue]{hyperref}
\usepackage{ulem}

\newcommand{\hoti}{Ho$_2$Ti$_2$O$_7$}

\newcommand{\HIO}{Ho$_2$Ir$_2$O$_7$}
\begin{document}

\title{Magnetic charge injection in spin ice: a new way to fragmentation}

\author{E. Lefran\c cois}
\altaffiliation[]{current address: Max Planck Institute for Solid State Research, Stuttgart, Germany}
\affiliation{Institut Laue Langevin, CS 20156, 38042 Grenoble, France}
\affiliation{Institut N\'eel, CNRS and Univ. Grenoble Alpes, 38042 Grenoble, France}
\author{V. Cathelin}
\affiliation{Institut N\'eel, CNRS and Univ. Grenoble Alpes, 38042 Grenoble, France}
\author{E. Lhotel}
\affiliation{Institut N\'eel, CNRS and Univ. Grenoble Alpes, 38042 Grenoble, France}
\author{J. Robert}
\affiliation{Institut N\'eel, CNRS and Univ. Grenoble Alpes, 38042 Grenoble, France}
\author{P. Lejay}
\affiliation{Institut N\'eel, CNRS and Univ. Grenoble Alpes, 38042 Grenoble, France}
\author{C. V. Colin}
\affiliation{Institut N\'eel, CNRS and Univ. Grenoble Alpes, 38042 Grenoble, France}
\author{B. Canals}
\affiliation{Institut N\'eel, CNRS and Univ. Grenoble Alpes, 38042 Grenoble, France}
\author{F. Damay}
\affiliation{Laboratoire L\'eon Brillouin, CEA, CNRS, Univ. Paris-Saclay , F-91191 Gif-sur-Yvette, France}
\author{J. Ollivier}
\affiliation{Institut Laue Langevin, CS 20156, 38042 Grenoble, France}
\author{B. F\aa k}
\affiliation{Institut Laue Langevin, CS 20156, 38042 Grenoble, France}
\author{L. C. Chapon}
\affiliation{Institut Laue Langevin, CS 20156, 38042 Grenoble, France}
\author{R. Ballou}
\affiliation{Institut N\'eel, CNRS and Univ. Grenoble Alpes, 38042 Grenoble, France}
\author{V. Simonet}
\affiliation{Institut N\'eel, CNRS and Univ. Grenoble Alpes, 38042 Grenoble, France}
\email[Corresponding author: ]{virginie.simonet@neel.cnrs.fr}

%\date{\today}

\maketitle

%%%%%%%%%%%%%%%%%%%%%%%%%%%%%%%%%%%%%%%%%
%
%%
%%{\color{red}\noindent  introduction: 150 mots - background - rationale - main results (here we show) - implications \\
%%corps du texte: 1500 mots \\
%% 3-5 Figures \\
%%30 ref max \\}
%
{\bf The complexity embedded in condensed matter fertilizes the discovery of new states of matter, enriched by ingredients like frustration. Illustrating examples in magnetic systems are Kitaev spin liquids \cite{Kitaev06}, skyrmions phases \cite{Rossler06}, or spin ices \cite{Harris97}. These unconventional ground states support exotic excitations, for example the magnetic charges in spin ices, also called monopoles \cite{Castelnovo08}. Beyond their discovery, an important challenge is to be able to control and manipulate them. Here, we propose a new mechanism to inject monopoles in a spin ice through a staggered magnetic field. We show theoretically, and demonstrate experimentally in the \HIO\ pyrochlore iridate, that it results in the stabilization of a monopole crystal, which exhibits magnetic fragmentation \cite{Bartlett14}. In this new state of matter, the magnetic moment fragments into an ordered part and a persistently fluctuating one \cite{Bartlett14,Petit16, Canals16, Paddison16}. Compared to conventional spin ices, the different nature of the excitations in this fragmented state opens the way to novel tunable field-induced and dynamical behaviors. } 

The spin ice state emerges in pyrochlore lattices of vertex sharing tetrahedra when the magnetic moments are subjected to an effective ferromagnetic interaction and are constrained along the local $\langle 111 \rangle$ directions joining the corners to the center of each tetrahedron (diagonals of the cubic crystal) \cite{Harris97}. It is, classically, a macroscopically degenerate ground state, in which the spins locally obey the ice-rule, with two spins pointing in and two spins pointing out of each tetrahedron (2i2o). Spin ice is the realization of a Coulomb phase \cite{Henley10}, since this local constraint, the ice-rule, can be mapped on an emergent divergence free field, which is actually the local magnetization. This results in the famous ``pinch point" pattern in the magnetic diffuse scattering \cite{Isakov04, Henley05, Fennell09}.
The spin ice elementary excitations, the magnetic monopoles, are obtained by flipping one spin at the center of a pair of tetrahedra. This produces two effective magnetic charges ($q=\pm1$), corresponding to 3 spins in--1 spin out (3i1o) and 1 spin in--3 spins out (1i3o) configurations \cite{Castelnovo08}, that can be deconfined.

\begin{figure*}
\includegraphics[width=17cm]{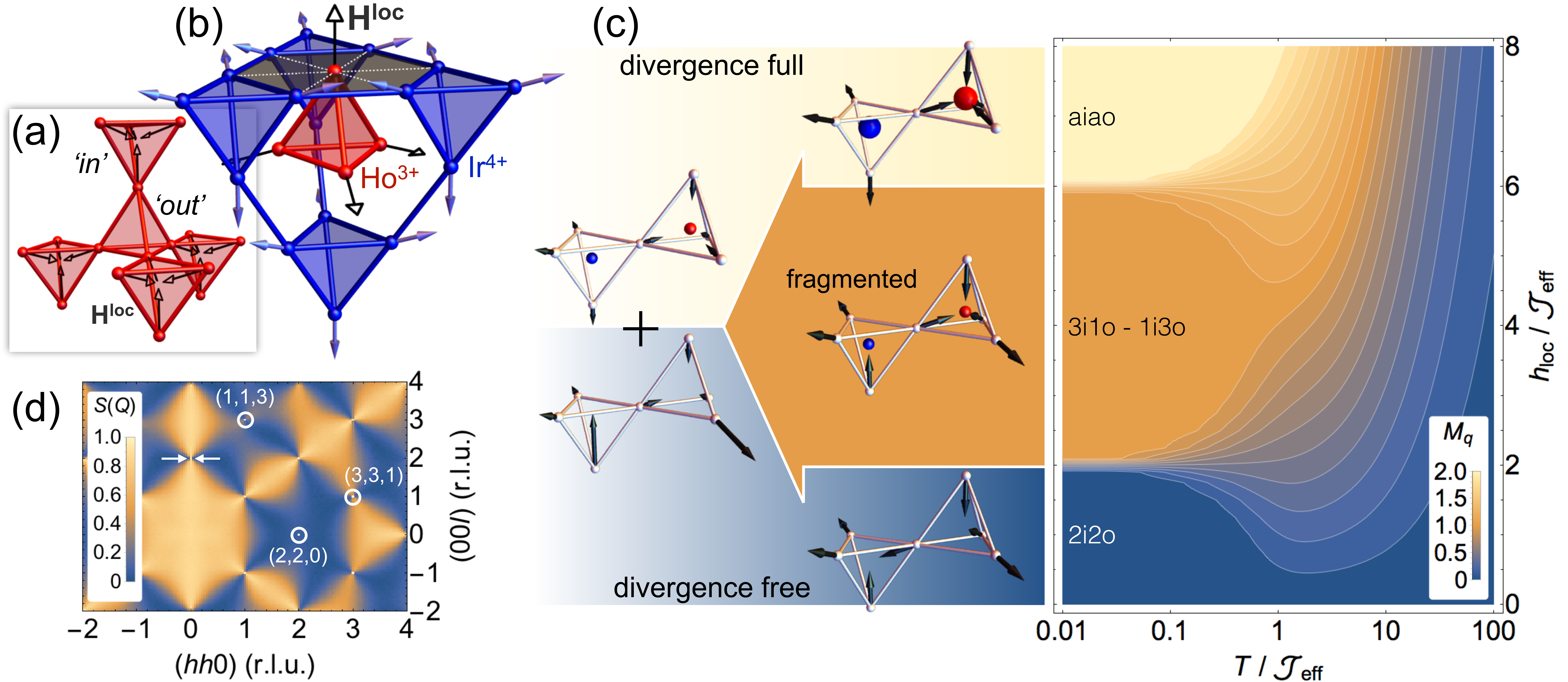}
\caption{\label{figstruct} {\bf Staggered field in a pyrochlore lattice and magnetic fragmentation}
(a) Pyrochlore lattice submitted to a $\langle 111 \rangle$ staggered field represented by the arrows on each site, with ``in'' and ``out'' tetrahedra when this field points either inward or outward of the tetrahedra. 
(b) Structure of the pyrochlore iridate Ho$_2$Ir$_2$O$_7$ with the Ho$^{3+}$ ions (red) and the Ir$^{4+}$ ions (blue), both occupying a pyrochlore lattice. Each rare-earth is surrounded by a hexagon of six Ir nearest neighbors. When the iridium lattice orders magnetically in the aiao phase, as shown on the blue lattice, a local magnetic field, perpendicular to the hexagon plane, hence aligned along the local $\langle 111 \rangle$ directions, is felt by the central rare-earth ion, as represented in (a). 
(c) Phase diagram as a function of reduced temperature $T/{\cal J}_{\rm eff}$ and local field $h_{\rm loc}/{\cal J}_{\rm eff}$, defined through the charge order parameter $M_q=\left \langle \left | \frac{1}{N}  \sum_{\alpha=1}^{N}  \Delta_{\alpha} q_{\alpha} \right| \right\rangle$ where $N$ is the number of tetrahedra and $\Delta_{\alpha}=+1$ (-1) for ``in" (``out") tetrahedra carrying a charge $q_{\alpha}$. As $h_{\rm loc}/{{\cal J}_{\rm eff}}$ increases, the ground state changes from the spin-ice ($M_q=0$, blue), to the aiao state ($M_q=2$, yellow), going through the charge crystal intermediate phase which fragments ($M_q=1$, orange).
The fragmentation can be described by the splitting of the pseudo spin $\sigma=\pm1$: the divergence free fragment on a tetrahedron can then be written as a state with four spins such that $\sigma_{\langle |q| \rangle=0}=(\pm 1/2, \pm 1/2, \pm1/2, \mp 3/2)$, while the divergence full fragment that carries the magnetic charge corresponds to an aiao configuration with half of the moment so that $\sigma_{\langle |q| \rangle=1}=(\pm 1/2, \pm 1/2, \pm1/2, \pm 1/2)$ \cite{Bartlett14}. All the spin and charge configurations are shown in the left panel of (c). 
(d) Magnetic scattering function $S(|\vec Q|)$, with $\vec Q$ the scattering vector, calculated in the fragmented phase. It shows a diffuse pattern with pinch points (evidenced by arrows) together with the magnetic Bragg peaks (highlighted by circles) characteristic of the aiao ordered state, but whose intensity (proportional to the magnetic moment squared) is a quarter of that expected when the whole magnetic moment is ordered.}
\end{figure*}

In the presence of dipolar interactions, and provided the charge density is large enough, the Coulomb like interaction between monopoles can stabilize a monopole crystal (staggered 3i1o and 1i3o configurations) in an underlying spin disordered manifold. This results in the fragmentation of the magnetization \cite{Bartlett14}, a state where a unique degree of freedom, the Ising magnetic moment, fragments thermodynamically into two parts, each part sustaining a different phase: an ordered antiferromagnetic phase, also interpreted as a crystal of magnetic charges (divergence full), and a disordered Coulomb phase (divergence free) with predominant ferromagnetic correlations \cite{Bartlett14}. 

A very appealing proposal is to generate this fragmentation by injecting magnetic charges in a controlled way, i.e. with a tunable parameter such as a magnetic field, without starting from this required state of monopole crystal, which imposes strong constraints on the Hamiltonian \cite{Bartlett14}. However, any external magnetic field does not act homogeneously on the magnetic moments of the unit cell, since the magnetic moments inside a tetrahedron point in different $\langle 111 \rangle$ directions. When the field is applied along one of the $\langle 111 \rangle$ directions, along which the pyrochlore lattice can be viewed as a stacking of triangular and kagome planes, only a ``partial" fragmentation occurs (within the kagome planes). This leads to the so called kagome ice phase \cite{Moessner03, Tabata06, Bartlett14} in which a charge crystal is stabilized in the kagome planes through a ``2 in--1 out, 1 in--2 out" rule. However, no magnetic charges are actually present in the tetrahedral units. 

We propose below a new approach to produce the fragmentation process in the whole pyrochlore lattice, even in absence of dipolar interactions, by considering a magnetic field exerted along the local directions of the magnetic moments. As depicted in figure \ref{figstruct}(a), if the sign of this field alternates, pointing inward for a given tetrahedron (called ``in") and outward for the neighboring tetrahedra (called ``out"), we show that this staggered field competes with the spin ice state. This results, over a large field range, in a fragmented ground state supporting unconventional excitations. 

Consider the nearest neighbor spin ice Hamiltonian in the presence of a local magnetic field:
\begin{equation}
\begin{aligned}
%\begin{split}
{\cal H} &= -{\cal J}\sum_{\langle i,j\rangle}{\bf{S}_i \cdot \bf{S}_j }- g \mu_0{\mu_{\rm B}} \sum_{j} {{\bf{H}^{loc}_i \cdot \bf{S}_i}} \\
&= {{\cal J}_{\rm eff}} \sum_{\langle i,j\rangle}{\sigma_i \sigma_j }-h_{\rm{loc}} \sum_{i} {\sigma_i}
\end{aligned}
% \end{split}
\label{eqH}
\end{equation} 
where $\bf{S_i}$ is the i$^{th}$ magnetic moment pointing along its local trigonal direction and interacting with its nearest neighbors via ${\cal J}$. $\sigma_i=\pm 1$ is the corresponding Ising pseudo-spin which is equal to +1 (-1) when the moment points inward (outward) the tetrahedron. ${{\cal J}_{\rm eff}}=-{\cal J}/3S^2>0$ is the effective nearest neighbor ferromagnetic interaction \cite{Jaubert11}. $\bf{H}^{loc}_i$ is the staggered magnetic field described above, aligned along the $\langle 111\rangle$ direction of the $i^{th}$ site, and $h_{\rm loc}=g \mu_0 {\mu_{\rm B}} H^{\rm loc}S$. 
% definition sigma_i 

Minimization of equation (\ref{eqH}) on a single tetrahedron, together with Monte Carlo simulations on a pyrochlore lattice (see figure \ref{figstruct}(c)), gives a succession of ground states with a different charge order parameter, depending on $h_{\rm{loc}}/{\cal J}_{\rm eff}$. At low fields, the ground state is the 2i2o spin ice state ($\langle |q| \rangle = 0$). In the opposite limit of large fields, the stabilized state is the ``all in--all out" (aiao) antiferromagnetic order (charge crystal with $\langle |q| \rangle = 2$), in which all the spins of a given tetrahedron point either inward or outward, following the staggered field. %(see figure \ref{figstruct}(a-b))}.

More exotic physics appear in the intermediate regime $2 < h_{\rm loc}/{\cal J}_{\rm eff} < 6$ (orange region in the phase diagram of figure \ref{figstruct}(c)): the competition between ${\cal J}_{\rm eff}$ and $h_{\rm loc}$ selects 3i1o and 1i3o configurations on the ``in" and ``out" tetrahedra respectively. It thus stabilizes a charge crystal, with $\langle |q| \rangle = 1$, and leads to the fragmentation process \cite{Bartlett14}, characterized by a staggered magnetization at half of the magnetic moment. Another important signature is the coexistence of diffuse scattering with a pinch point pattern and aiao Bragg peaks in the computed scattering function (see figure \ref{figstruct}(d)). Here, the fragmented state does not result from the presence of the long-range monopole Coulomb energy but from the injection at equilibrium of magnetic charges into a spin ice state by a staggered magnetic field.

A prominent system to realize this model is the pyrochlore iridate family, of formula $R_2$Ir$_2$O$_7$ in which $R$ is a rare-earth element. In these compounds, the iridium and rare-earth ions lie on two interpenetrated pyrochlore lattices (see figure \ref{figstruct}(b)). It was shown in pyrochlore iridates where $R$=Nd, Eu, Tb, that the iridium sublattice orders in an aiao magnetic arrangement \cite{Guo16, Sagayama13, Lefrancois15}, simultaneously with a metal insulator transition \cite{Matsuhira11} ($T_{\rm MI}=30-140$~K).
This arrangement generates a molecular field oriented along the local $\langle 111 \rangle$ directions on the rare-earth sublattice \cite{Lefrancois15}, and thus plays exactly the same role as our staggered magnetic field $h_{\rm loc}$ in equation (\ref{eqH}). The fragmentation process described above can thus be realized in the pyrochlore iridates, provided the rare-earth ions have a local easy axis anisotropy along the $\langle 111 \rangle$ directions and interact through a ferromagnetic interaction. %, large enough to compete with the Ir molecular field. 
This is the case of Ho and Dy ions in pyrochlore compounds with non magnetic ions instead of Ir$^{4+}$, and where spin ice physics is observed at Kelvin temperatures \cite{Harris97, Ramirez99, Kadowaki02,Zhou11,Hallas16}. %Ho and Dy pyrochlores are the archetype of spin ice systems when the Ir$^{4+}$ ion is replaced by a non-magnetic ion \cite{Harris97, Ramirez99, Kadowaki02,Zhou11,Hallas16}: they have a strong Ising anisotropy and exhibit an effective ferromagnetic interaction in the kelvin range, resulting from exchange and dipolar interactions. Pyrochlore iridates with these rare-earth ions are thus suitable candidates for field-induced magnetic fragmentation. 
In the present article, we focus on the Ho pyrochlore iridate.   

\begin{figure}
\includegraphics[width=9cm]{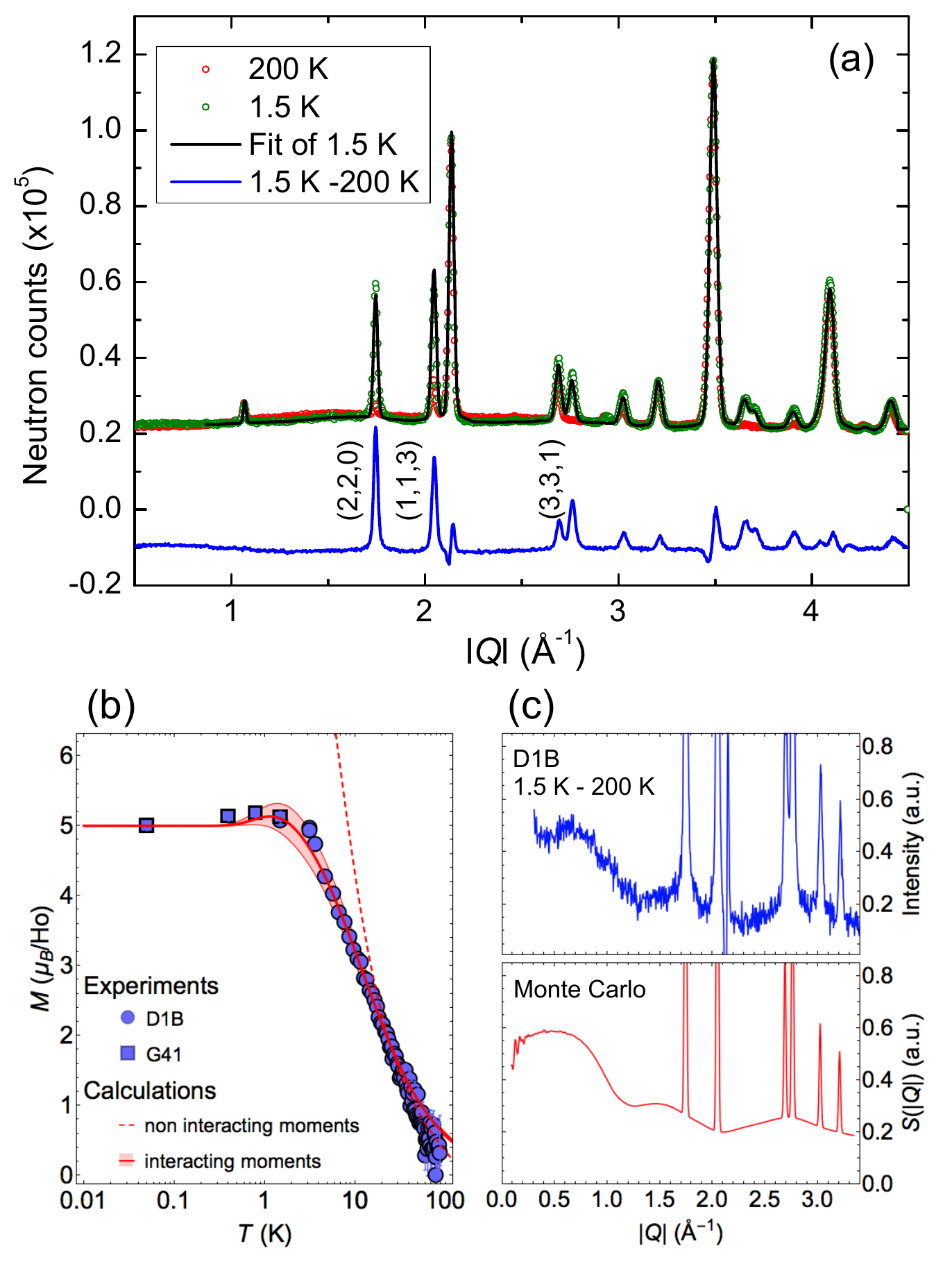}
\caption{\label{figneut} {\bf Magnetic ordering and diffuse scattering in \HIO}
(a) Neutron diffractograms recorded at 200 K (red dots) and 1.5 K (green dots) showing the rise of magnetic Bragg peaks. The 1.5 K Rietveld refinement (black line) using a ${\bf k}={\bf 0}$ propagation vector yields an aiao magnetic order for the holmium sublattice. The difference between the 1.5 and 200~K data (blue line, shifted for clarity) enhances the magnetic Bragg peaks. % especially at the (220) and (311) positions. 
(b) Temperature evolution of the refined Ho$^{3+}$ magnetic moment, in a semi-logarithmic scale, measured with two diffractometers (D1B and G4.1). The red dashed curve is the calculated ordered magnetic moment induced by the temperature dependent molecular field created by the Ir magnetic ordering (see Supplementary Information). Below 20 K, the calculated and experimental curves depart from each other due to the presence of Ho-Ho magnetic interactions. The red full curve is obtained from Monte Carlo calculations (equation (\ref{eqH})) with ${\cal J}_{\rm eff}=1.4$~K and $h_{\rm loc}/{\cal J}_{\rm eff}=4.5$, the colored area representing the calculated values compatible with the experimental error bars. These calculations allow to account for the saturation of the ordered moment below 2~K due to fragmentation. The departure above 40 K between the calculated and experimental curves is attributed to the decrease of the Ir molecular field. (c) Evidence for diffuse scattering at 1.5 K (top), from the difference between the 1.5 and 200 K diffractograms (the negative intensity is due to an imperfect subtraction of a strong nuclear peak caused by the lattice parameter variation). It is compared to Monte-Carlo calculations (bottom) at $T=0$ with %the above determined parameter 
$h_{\rm loc}/{\cal J}_{\rm eff}=4.5$.}
\end{figure}

Our measurements were performed on a high quality polycrystalline sample of \HIO. As previously reported \cite{Matsuhira11}, the magnetization exhibits a zero field cooled - field cooled (ZFC-FC) bifurcation at $T_{\rm MI}$=140 K pointing out the magnetic ordering of the Ir sublattice (see Supplementary Information). Below this temperature, the iridium sublattice starts to produce a molecular field on the holmium sites, which manifests as the rise of magnetic Bragg peaks below about 80~K in the neutron diffraction pattern. Due to the very weak moment of the iridium, these Bragg peaks are mainly characteristic of an aiao magnetic ordering of the holmium sublattice (see figure \ref{figneut}(a)). This is the ordering expected in our model when the temperature and the staggered magnetic field are large with respect to the effective interaction. From the temperature evolution of the refined Ho$^{3+}$ ordered magnetic moment, strongly increasing when decreasing the temperature, we could extract the value of the molecular field $H^{\rm loc}$ estimated to $0.725 \pm 0.05$~T below 40~K (see figure \ref{figneut}(b)). This was achieved following the procedure of reference \onlinecite{Lefrancois15} and using the crystal electric field parameters of the Ho$^{3+}$ ion deduced from inelastic neutron experiments (see Supplementary Information).

When decreasing further the temperature, the Ho-Ho magnetic interactions start to become relevant. The ordered moment finally saturates below 2 K, with a slight overshoot, ending with a value of $5 \pm 0.06~\mu_{\rm B}$ at 50~mK. This value is only half the magnetic moment of the Ho$^{3+}$ ions in their ground state doublet (see Supplementary Information). A diffuse scattering signal is actually present and accounts for the missing moment, as seen in figure \ref{figneut}(c). It has broad maxima at  $Q \sim0.7$, 1.6 and 2.7~\AA$^{-1}$, evidencing additional spin correlations characteristic of the spin ice diffuse scattering in polycrystalline samples \cite{Kadowaki02, Hallas16}. This coexistence of a spin ice diffuse scattering with an aiao ordering, whose ordered magnetic moment is half of the total moment, is remarkably well reproduced in the calculations (see figure \ref{figneut}(c)) and is the signature that a fragmentation process occurs in \HIO\ below 2~K. Finally, the calculated temperature dependence of the ordered moment using equation (\ref{eqH}) gives ${\cal J}_{\rm eff}=1.4 \pm 0.1$~K and $h_{\rm loc}/{\cal J}_{\rm eff}=4.5 \pm 0.25$ (see figure \ref{figneut}(b)), which places \HIO\ deep into the fragmentation regime according to figure \ref{figstruct}(c). It corresponds to a local field $\mu_0 H^{\rm loc}=0.94 \pm 0.02$ T, in the same range as the molecular field estimated from the high temperature behavior. 

\begin{figure}
\includegraphics[width=8.5cm]{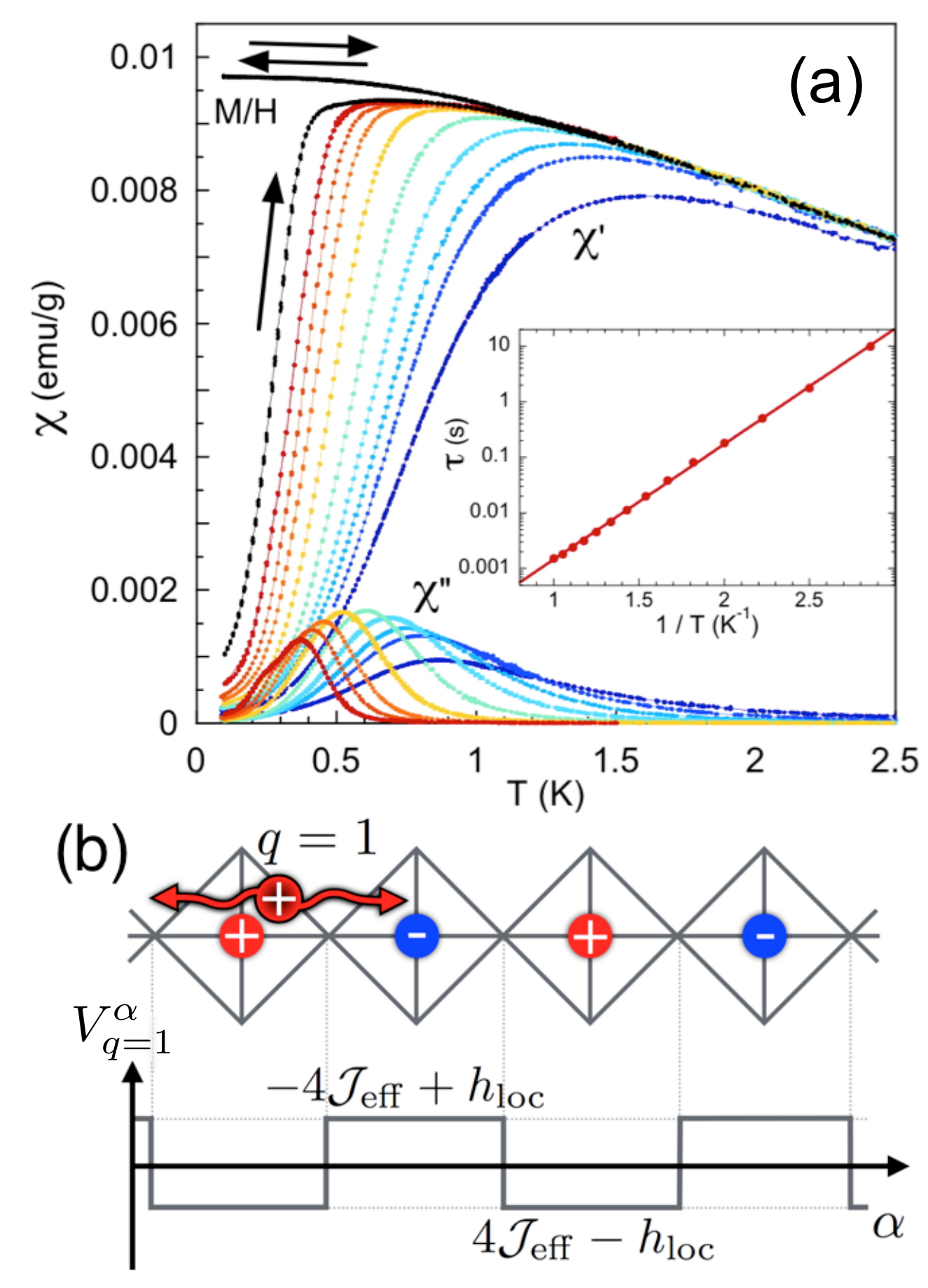}
\caption{
\label{figXac} {\bf Dynamical properties and diffusion of the excitations}
(a) Low temperature dependence of the ZFC-FC magnetization performed in 50 Oe below 4 K. The two curves depart below about 1.5 K. On the same graph are shown the real part $\chi'$ and imaginary part $\chi''$ of the ac susceptibility vs $T$ at several frequencies $f$ between 0.011 and 570~Hz. Inset: $\tau$ vs $1/T$ where $\tau=1/2\pi f_{\rm peak}$ and $f_{\rm peak}$ is the $\chi''$ maximum of $\chi''(f)$ measured at constant temperature $T$ (see Supplementary Information). The line is a fit to an Arrhenius law $\tau=\tau_0 \exp(E/T)$ where $\tau_0= (1.2 \pm 0.2) \times 10^{-5}$~s and $E=4.8 \pm 0.1$ K. 
(b) Top: two dimensional representation of a string of tetrahedra on which an elementary excitation carrying a charge $q=+1$ (red disc with black border) diffuses in the fragmented regime. 
Bottom: spatial periodic potential $V_q^\alpha$ felt by the diffusing charge and induced by the underlying static charge crystal, where $\alpha$ refers to the tetrahedron type: $\alpha=$``in" carries an ordered charge $q_{o}^{in}=+1$ (red) and $\alpha=$``out" carries $q_{o}^{out}=-1$ (blue). 
%This potential writes $V_\alpha (q) = (4{\cal J}_{\mathrm{eff}} - h_{\mathrm{loc}}) \, q_{o}^\alpha \, q$. The energy of a single excitation is then given by $E^\alpha(q) = 2{\cal J}_{\mathrm{eff}} + V_\alpha(q)$.
}
\end{figure}

We have thus demonstrated that the thermodynamical fragmentation of the magnetic moment under the influence of a local staggered field predicted in our model appropriately describes the unexpected properties of \HIO. 
In the following, we probe the macroscopic magnetic properties of \HIO\  to explore the excitations emerging from this new fragmented state. 

In the fragmented regime, typically below 2~K, the ac susceptibility displays a frequency dependence, which can be described by a thermally activated process above an energy barrier $E\approx 4.8$~K (see figure \ref{figXac}(a)). It is associated with a freezing in the ZFC-FC magnetization below 1 K, where a strong irreversibility is observed. Similar slow dynamics are  observed in spin ice materials \cite{Matsuhira11b, Quilliam11}, where they are induced by the diffusion of monopoles, with a relaxation timescale inversely proportional to the monopole density \cite{Jaubert09, Jaubert11, Castelnovo11}. In the present case, the dynamics can also be understood by the diffusion of the fractional excitations, of energy $2{\cal J}_{\mathrm{eff}}$ and charge $q = \pm 1$, emerging from the fragmented Coulomb phase. But a marked difference with the spin ice is that these excitations are diffusing in a periodic potential $V_q^\alpha$ that depends on the local field. It is created by the underlying charge crystal fragment through a local interaction between the diffusive charge and the ordered charges. 
The energy of a single excitation is given by
\begin{equation}
E^\alpha_q = 2{\cal J}_{\mathrm{eff}} + V_q^\alpha,
\label{eqEq}
\end{equation} 
with $V_q^\alpha = (4{\cal J}_{\mathrm{eff}} - h_{\mathrm{loc}}) \, q_{o}^\alpha \, q$, and where $q_{o}^\alpha = - 1$,  $+ 1$ is the ordered charge on a tetrahedron with $\alpha =$ ``out", ``in" (see fig. \ref{figXac}(b)).
For $h_{\rm loc} / {\cal J}_{\rm eff}=4.5$, corresponding to the \HIO\ case, the periodic potential is alternatively repulsive (attractive) for a diffusing charge interacting with an ordered charge of opposite (same) sign (see figure \ref{figXac}(b)). The relaxation timescale is then given by the highest energy barrier $E=h_{\rm loc} - 2{\cal J}_{\rm eff} \simeq 3.5$~K, rather close to the experimental value. Long range dipolar interactions \cite{Jaubert11}, as well as more complex diffusion mechanisms (see Supplementary Information), could significantly affect the relaxation timescale and improve the agreement with experiments.

\begin{figure}
\includegraphics[width=8.5cm]{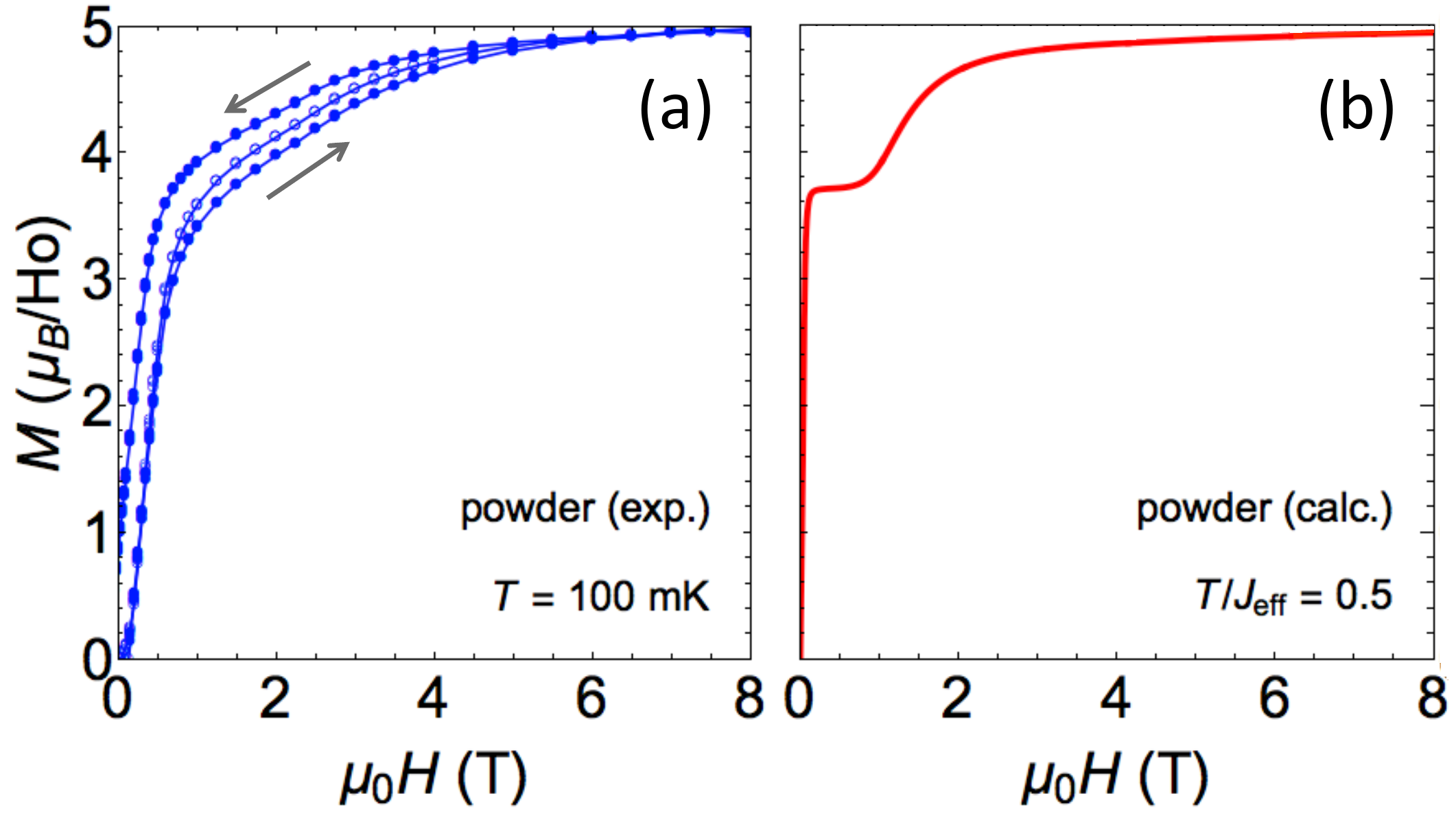}
\caption{\label{figMH} {\bf Magnetization, measurements and calculations} 
(a) $M$ vs $H$ measured at 110~mK. The empty symbols show the first magnetization curve and the filled symbols the magnetization curves obtained afterwards by ramping the field up and down. The small hysteresis opening is due to the existence of slow dynamics at low temperature. (b) Calculated $M$ vs $H$ for $h_{\rm loc}/{{\cal J}_{\rm eff}}=4.5$ and $T/{\cal J}_{\rm eff}=0.5$, performing a powder averaging. The anomalies, mimicking the experimental curve, feature the existence of an exotic field-induced behavior exhibiting magnetization plateaus (see Supplemental Information). 
%unit\'e H/Jeff: en fait c'est $g \mu_B S H(T)/{\cal J}_{\rm eff} k_B$ qui est trac\'e.
}
\end{figure}

In addition to the peculiar excitations hosted by this fragmented state, the application of an external uniform magnetic field is expected to promote exotic phases. In \HIO, isothermal magnetization curves at very low temperature do exhibit an unconventional behavior:  the magnetization first rises abruptly to reach a plateau-like feature at $\approx 3.5~\mu_{\rm B}$/Ho, before increasing again above 2 T up to 5 $\mu_{\rm B}$ at 8 T (which is the expected value for Ising Ho moments along the $\langle 111\rangle$ directions \cite{Bramwell00}). This fine structure of the magnetic isotherm is qualitatively reproduced by Monte-Carlo calculations in the fragmented state using a powder average (see figure \ref{figMH}(b)). It actually reflects the rich plateau physics emerging from the fragmented ground state. Several field-induced phases are stabilized depending on the orientation of the external magnetic field, for example successive fragmented kagome ice states (see Supplementary Information). Further experimental and theoretical works are required to probe these exotic states in more details%, and to compare them with the field-induced behavior of conventional spin ices  
 \cite{Moessner03, Tabata06,Jaubert08, Fennell07}. 

The introduction of a staggered magnetic field in the nearest neighbor spin ice Hamiltonian on the pyrochlore lattice thus allows, through magnetic charge injection, to stabilize a fragmented state, and offers a rich playground to study the various facets of spin ice and monopoles physics in a renewed perspective. We have shown that, in the pyrochlore iridate \HIO, the ratio between the rare earth interactions and the staggered field induced by the iridium moment places the system deeply in the fragmented phase. Experimentally, it would then be very appealing to explore the full fragmentation phase diagram and the associated exotic dynamics by varying the ratio between rare earth interactions and the staggered local field. This can be done by changing the rare-earth element. Preliminary measurements on the Dy pyrochlore iridate suggest that fragmentation also occurs in this compound, and hopefully, this system will be located at a different $h_{\rm loc}/{\cal J}_{\rm eff}$. This phase diagram mapping should also be achieved more systematically by applying an external isostatic pressure. Finally, the model we have developed is perfectly equivalent to the case of antiferromagnetically coupled uniaxial Ising spins on a pyrochlore lattice submitted to a uniform external field, which can be easily varied. It thus opens the route to the tuning of fragmentation in new systems. 

%%%%%%%%%%%%%%%%%%%%%%%%%%%%%%%%%%%%%%%%
\section*{Methods}
%%%%%%%%%%%%%%%%%%%%%%%%%%%%%%%%%%%%%%%%
{\it Calculations - } %Thermodynamical quantities as well as structure factors have been calculated through Monte Carlo simulations of the spin model described by equation \ref{eqH}. The single spin-flip Metropolis update, which reproduces the local dynamics relevant for materials such as Ho$_2$Ir$_2$O$_7$, has been combined to a cluster algorithm preventing the system from being trapped into a small region of the configuration space at low temperature. 
%(for details about the methods, see Supplementary Information).
%The simulations were performed on $16 \times L^3$ lattice sites with $L=8,12, 24$ depending on the calculated quantities, and with periodic boundary conditions. $10^4$ hybrid Monte Carlo steps (HMCS) have been used for thermalization, and measurements are computed over $N$ HMCS, where $N$ is adapted to ensure stochastic decorrelation between measurements (typically, from $N=10^3$ at high temperature to $10^6$ at low temperature).
Thermodynamical quantities as well as structure factors have been calculated through Monte Carlo simulations of the spin model described by equation \ref{eqH}. 
A single spin-flip Metropolis update has been combined to a loop algorithm preventing the system from being trapped into a small region of the configuration 
space at low temperature. The simulations were performed on $16 \times L^3$ lattice sites with $L=8,12$ (figures \ref{figstruct}(d) and \ref{figMH}(b)) and $24$ (figures \ref{figstruct}(c) and \ref{figneut}(c) bottom) with periodic boundary conditions. $10^4$ hybrid Monte Carlo steps were used for thermalization, while the measurements were computed over $N$ steps, where $N$ is adapted to ensure stochastic decorrelation between measurements (typically, from $N=10^4$ at high temperature to $10^7$ at low temperature). The powder magnetization of figure~\ref{figMH}(b), averaged over 800 random field directions, has been computed at sufficiently high temperature to preserve the efficiency of the used algorithm.

{\it Synthesis - } Polycrystalline samples of R$_2$Ir$_2$O$_7$, with R = Ho$^{3+}$ (4f$^{10}$, $S$=2, $L$=6, $J$=8, $g_J$=5/4, $g_JJ$=10 $\mu_B$) of high quality were synthesized by a mineralization process. The starting reagents R$_2$O$_3$ and IrO$_2$ were mixed together with a small amount of KF flux and pressed into a pellet. After being placed in a Pt crucible, they were submitted to a heat treatment under air atmosphere: 200$^{\circ}$C/h until a plateau of 120 h at 1110$^{\circ}$C and a decrease to room temperature at 200$^{\circ}$C/h. The pyrochlore iridates crystallize in the $Fd\bar3m$ cubic space group, with the O occupying the 48$f$ and 8$b$ Wyckoff positions, the rare-earth and the Ir occupying the 16$d$ and 16$c$ positions respectively. The structure and quality of the samples were checked by X-ray diffraction, with no traces of parasitic phases. The lattice parameter and the $x$ coordinate of the 48$f$ O were found at room temperature equal to 10.1790(2) \AA\ and 0.343(2). %for the Ho compound. %and equal to \textcolor{red}{10.2378(5)} \AA\ and \textcolor{red}{0.35(2)} for the Dy compound. 

{\it Neutron diffraction - } The CRG-D1B diffractometer at the ILL (wavelength $\lambda$=2.52 \AA) was used in the $1.5-300$~K temperature range and the G4.1 diffractometer at the LLB ($\lambda$=2.427 \AA) was used for the lower temperature range. It is equipped with a Cryoconcept-France HD dilution refrigerator (100µW@100mK), and fitted with vanadium shields. The sample was put in a vanadium cell in 14 bars of He gaz. %For \DIO, due to the strong absorption of the Dy, the hot neutron D4c diffractometer was preferred with $\lambda=0.5$ \AA, in  a cryostat spanning the 2-300 K temperature range.
 Rietveld refinements were carried out with the FULLPROF program \cite{Rodriguez-Carvajal93}. 

{\it Magnetization measurements - } Very low temperature magnetization measurements were performed on SQUID magnetometers equipped with a dilution refrigerator developed at the Institut N\'eel \cite{Paulsen01}. The powder samples were crushed and mixed with Apiezon grease in a copper pouch to ensure a good thermal contact between the copper sample holder and the sample. Ac susceptibility measurements were performed with an ac field of 1 or 2.34 Oe. Demagnetization corrections were performed assuming an average demagnetization factor of $0.1$ (cgs units) estimated from the shape of the copper pouch.  

%%%%%%%%%%%%%%%%%%%%%%%%%%%%%%%%%%%%%%%
\section*{Acknowledgments}
%%%%%%%%%%%%%%%%%%%%%%%%%%%%%%%%%%%%%%%%
We acknowledge A. Hadj-Azzem and J. Balay for their help in the compound synthesis. 
We thank C. Paulsen for allowing us to use his SQUID dilution magnetometers and P.C.W. Holdsworth and S. Petit for fruitful discussions.
E. Lh and V. C. acknowledge financial support from ANR, France, Grant No. ANR-15-CE30-0004.

%%%%%%%%%%%%%%%%%%%%%%%%%%%%%%%%%%%%%%%%
%Bibliography
%%%%%%%%%%%%%%%%%%%%%%%%%%%%%%%%%%%%%%%%
%\bibliography{../Biblio_ErTb2Ir2O7}
%

%%%%%%%%%% Merge with supplemental materials %%%%%%%%%%
\widetext
\clearpage
\renewcommand{\thefigure}{S\arabic{figure}}
\renewcommand{\theequation}{S\arabic{equation}}
\renewcommand{\thesection}{\Alph{section}}
%\makeatletter
%\renewcommand{\@biblabel}[1]{\quad S#1. }
%\makeatother
\begin{center}
\textbf{\large Supplementary Information: \\Magnetic charge injection in spin ice: a new way to fragmentation}
\end{center}

\twocolumngrid

\maketitle

%%%%%%%%%%%%%%%%%%%%%%%%%%%%%%%%%%%%%%%%
%\section{abstract}
%%%%%%%%%%%%%%%%%%%%%%%%%%%%%%%%%%%%%%%%

%%%%%%%%%%%%%%%%%%%%%%%%%%%%%%%%%%%%%%%%
\section{Magnetization measurements above 2 K}
%%%%%%%%%%%%%%%%%%%%%%%%%%%%%%%%%%%%%%%%

The temperature and field dependence of the magnetization ($M$) of the \HIO\ powder sample were measured using Quantum Design VSM and MPMS\textsuperscript{\textregistered} SQUID magnetometers down to 2 K. 

\begin{figure}
\includegraphics[width=8cm]{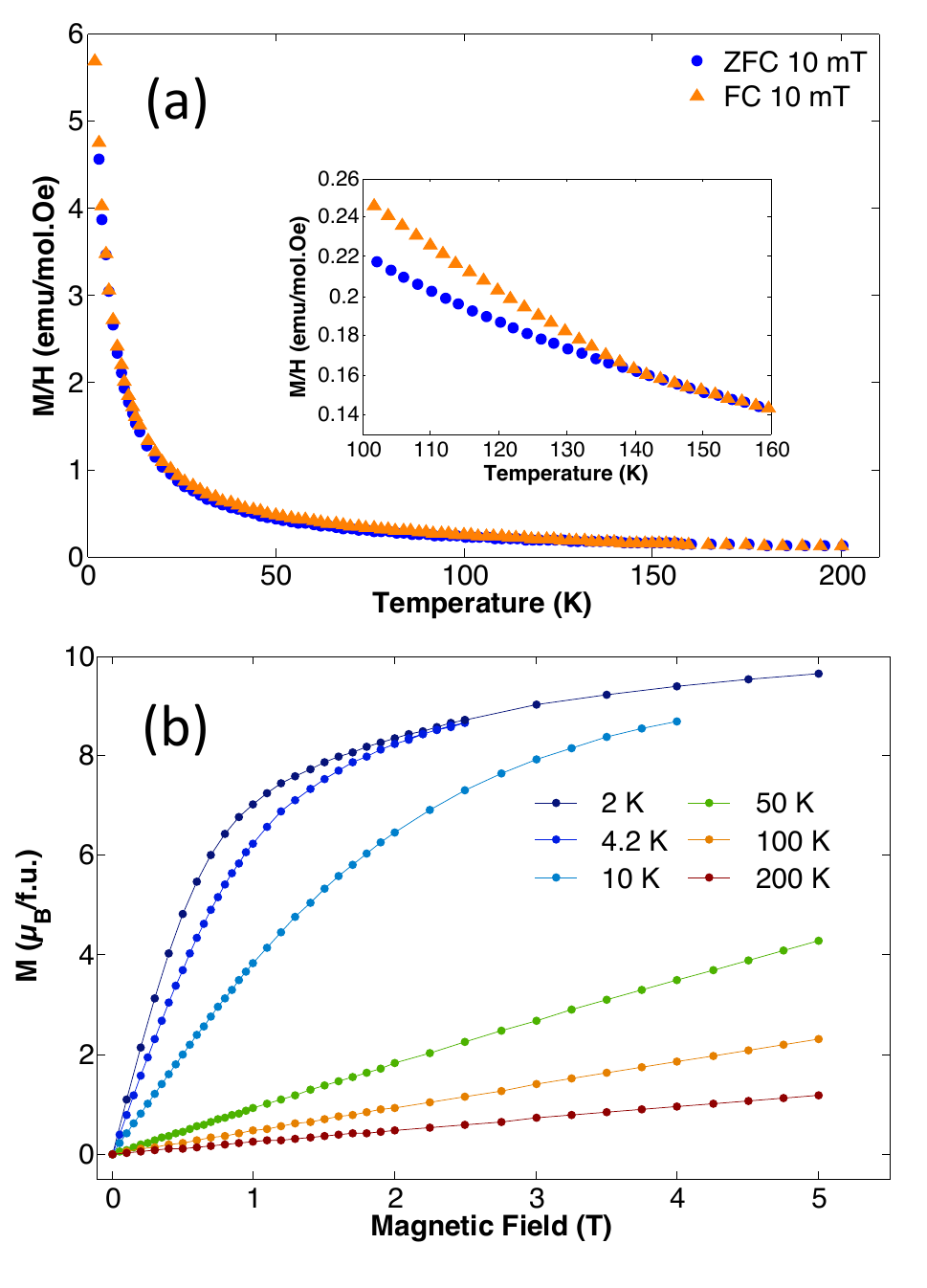}
\caption{\label{figM} (a) $M/H$ versus $T$ for \HIO\ measured after a ZFC (in blue) and a FC (in orange) procedure in 0.01 T. Inset: zoom of the ZFC-FC opening occurring around 140 K. (b) $M$ versus $H$ for \HIO\ measured at different temperatures. }
\end{figure}

The magnetization versus temperature exhibits a  Zero Field Cooled (ZFC) - Field Cooled (FC) bifurcation at T$_{MI}$=140 K pointing out the ordering of the Ir sublattice concomitant with the metal-insulator transition (see figure \ref{figM}(a)). The magnetic isotherms, linear at high temperature, are approaching a plateau of $\approx$5 $\mu_{\rm B}$/Ho at 2 K, which reflects the strong Ising anisotropy of Ho$^{3+}$ (see figure \ref{figM}(b)).

%%%%%%%%%%%%%%%%%%%%%%%%%%%%%%%%%%%%%%%%
\section{Inelastic neutron scattering}
%%%%%%%%%%%%%%%%%%%%%%%%%%%%%%%%%%%%%%%%

The magnetic excitations in \HIO\ were studied by inelastic neutron scattering (INS) on the same polycrystalline sample as the one used for neutron diffraction and magnetometry measurements. The experiments were carried out at the Institut Laue-Langevin on the IN6 and IN4 neutron spectrometers. 

\subsection{Experiments and results}

On the IN6 cold neutron time-of-flight spectrometer, the incoming neutron wavelength (energy) was 5.1~\AA~(3.145~meV). The energy resolution for elastic scattering was $\delta E \approx$ 70~$\mu$eV. No signal is detected at the lowest temperature $T = 1.5$~K in the accessible range of energy transfer ($\omega <$~2~meV).
Non dispersive signals become visible in the neutron energy gain channel when the temperature is increased, which were ascribed, in the light of the data obtained on IN4, to transitions between crystalline electric field (CEF) levels of the Ho$^{3+}$ ions.

\begin{figure}[h]
    \centering
    \includegraphics[width=0.48\textwidth]{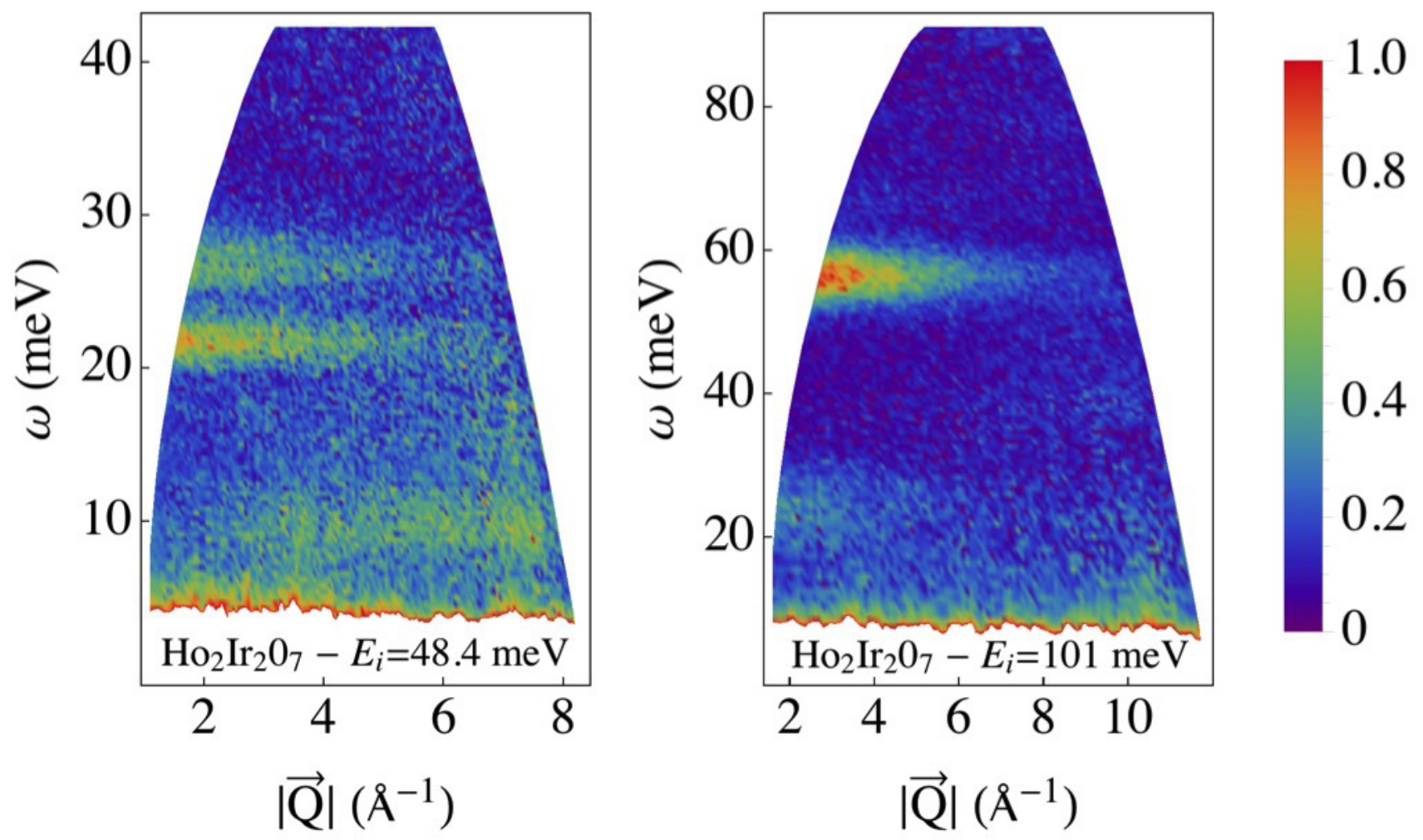}
    \caption{Scattering function $S(|\vec{Q}|, \omega)$ measured at 1.5~K for incoming neutron energy $E_i$ = 48.4~meV (left) and $E_i$ = 101~meV (right). The data are normalized with respect to the maximum of $S(|\vec{Q}|, \omega)$ in the inelastic region ($\omega \geq 1.5~Ê\delta$E).}     \label{fig:ExpSQw}
\end{figure}

On the IN4 thermal neutron time-of-flight spectrometer, the incoming neutron wavelengths (energies) used were 2.44~\AA~(13.74~meV),  1.3~\AA~(48.4~meV) and 0.9~\AA~(101~meV). The energy resolution for elastic scattering in the respective settings was $\delta E \approx 1$, 3.75 and 9.5~meV. The scattering function $S(|\vec{Q}|, \omega)$ measured at 1.5~K for the accessible scattering vector $\vec{Q}$ and energy transfer $\omega$ are displayed in figure~\ref{fig:ExpSQw}. 
\begin{figure}[t]
    \centering 
    \includegraphics[width=0.48\textwidth]{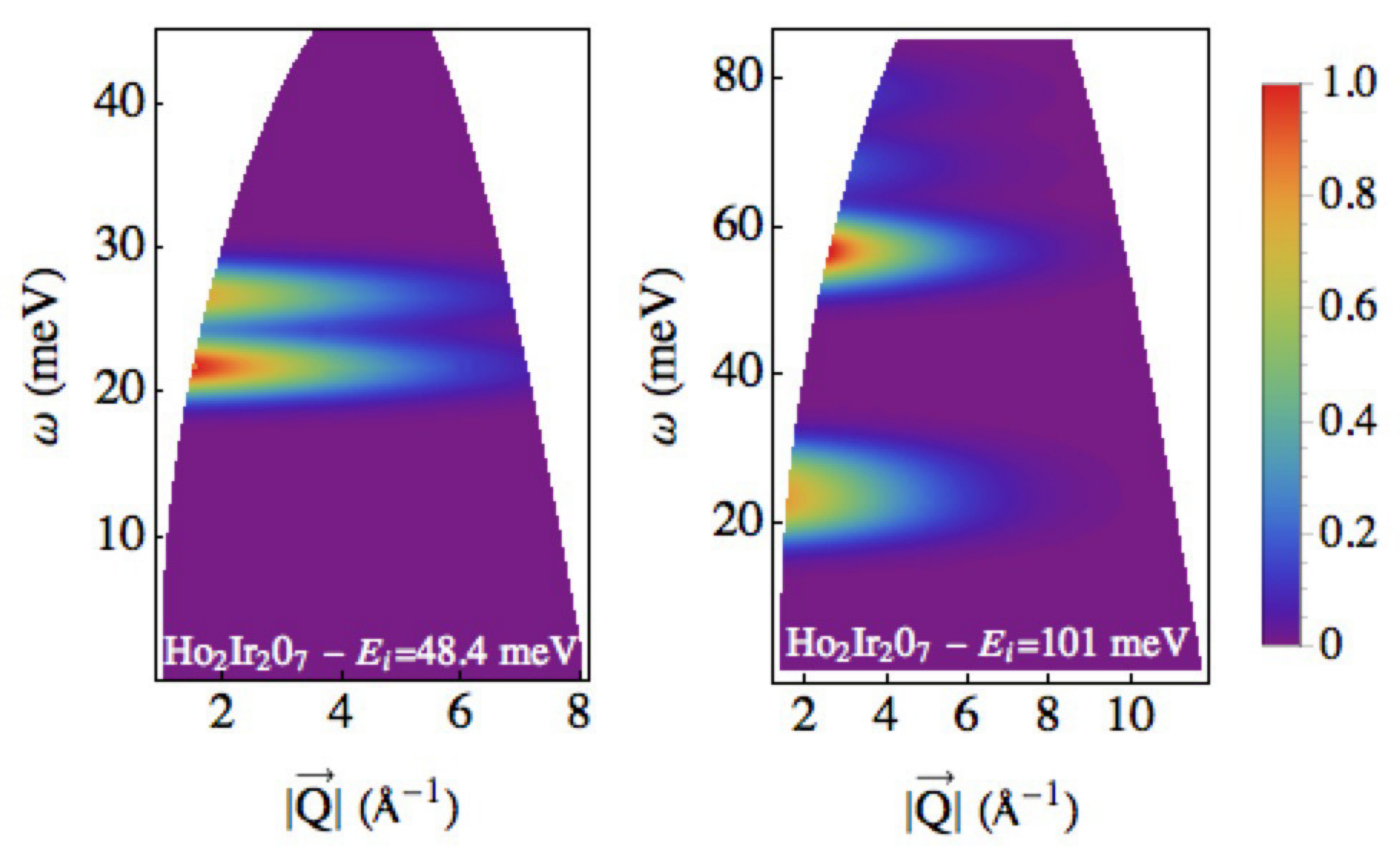}
    \includegraphics[width=0.28\textwidth]{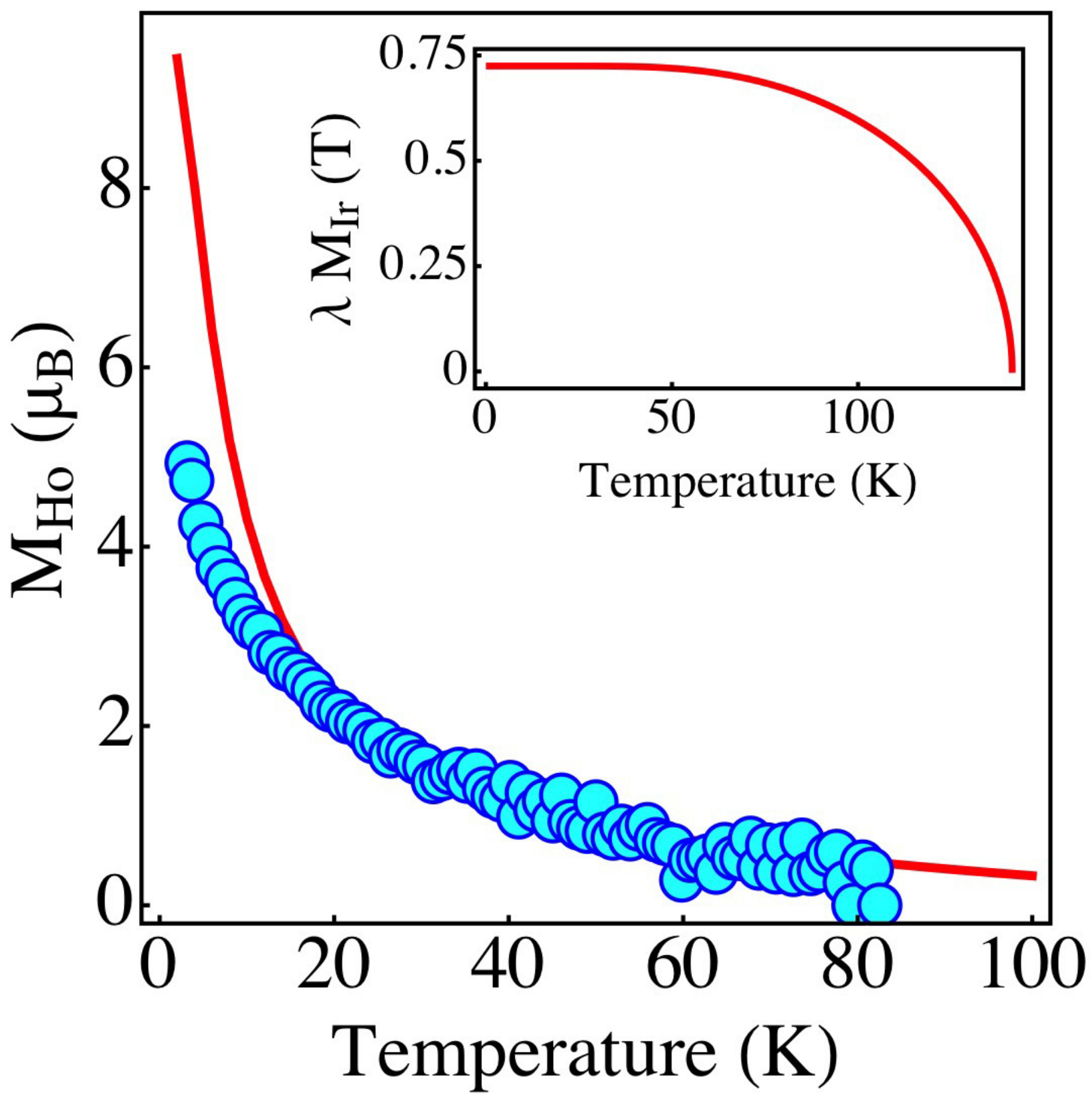}
    \caption{ Top panel: Scattering function $S(|\vec{Q}|, \omega)$, normalised by its maximum, calculated at 1.5~K in the neutron energy loss channel for incoming neutron energy $E_i = 48.4$~meV (left) and $E_i = 101$~meV (right). Bottom panel: Experimental (cyan disks) and calculated (red line) thermal variation of the ordered magnetic moment $M_{\rm Ho}$ of the Ho$^{3+}$ ions induced by the Ir-Ho molecular field. The inset shows the calculated thermal variation of the Ir-Ho molecular field $\lambda \vec{M}_{\rm Ir}(T)$ in Tesla units.}
     \label{fig:ExpSQwC}
\end{figure}
Three main signals are essentially observed, at energy transfer $\omega$ around 21,  25 and 56~meV. Their intensity decreases when $|\vec{Q}|$ increases, roughly like the square of the magnetic form factor calculated from the radial component of the wavefunction of the 4f$^{10}$ electronic configuration of the Ho$^{3+}$ ions \cite{ACL+06SM}. No dispersion is detected within the experimental accuracy. This suggests on one hand that these signals arise from CEF excitations and on the other hand that the Ir-Ho and Ho-Ho exchange energies that govern the collective magnetic behavior of \HIO\ at low temperature are smaller than the energy resolution in the experiment. All other signals are ascribed to structural dynamics, being observed at high $|\vec{Q}|$ or increasing with $|\vec{Q}|$. A weak signal was noticed at $\omega$ around 4-5~meV but a careful examination of its $|\vec{Q}|$ variation indicated that it cannot be ascribed to a CEF excitation. It can no more be ascribed to a split of the ground doublet by exchange fields. A quantitative estimate indeed shows that this would lead to detectable dispersion on the other excitations and to a thermal variation of the Ho$^{3+}$ magnetic moments very different from the one experimentally deduced from the neutron diffraction experiments.

\subsection {CEF analysis and determination of the Ir-Ho molecular exchange field}

\subsubsection{CEF Hamiltonian}
The CEF Hamiltonian for f electrons in the D$_{3d} (\bar{3}m)$ point group symmetry of the 16d site occupied by the Ho$^{3+}$ ions in the \HIO\ crystal writes 
\begin{eqnarray}
\mathcal{H}_{\rm CEF} & = & B_2^0\mathbf{C}_2^0 + \nonumber \\
 &    &  B_4^0\mathbf{C}_4^0 + B_4^3(\mathbf{C}_4^3 - \mathbf{C}_4^{-3}) +  \\
 &    & B_6^0\mathbf{C}_6^0 +  B_6^3(\mathbf{C}_6^3 - \mathbf{C}_6^{-3}) + B_6^6(\mathbf{C}_6^6 - \mathbf{C}_6^{-6}) \nonumber 
\end{eqnarray}
when the quantization axis is chosen along the local 3-fold axis. The $\mathbf{C}_k^q$
stand for Wybourne operators that, in a spatial rotation, transform like the spherical harmonics  $Y_k^q$. The $B_k^q$ are the (real) CEF parameters. They can be extracted from the experimentally recorded INS spectra by first diagonalizing $\mathcal{H}_{\rm CEF}$ on the ground multiplet $^{5}$I$_8$ ($L=6$, $S=2$, $J=8$) of the Ho$^{3+}$ ions, then by calculating from the obtained eigenvalues $\epsilon_n$ and eigenstates $|n\rangle$ the orientation averaged scattering function
\begin{equation}
S(|\vec{Q}|, \omega) = f(|\vec{Q}|)^2 \sum_{n,m}\frac{e^{-\beta\epsilon_n}}{Z}\frac{2}{3} |\langle n|\vec{\mathbf{J}}|m \rangle|^2 \mathcal{R}(\omega - \omega_{n \rightarrow m})
\end{equation}
where $\beta=\frac{1}{k_BT}$ is the temperature factor, $Z=\sum_n e^{-\beta\epsilon_n}$ is the partition function, $\vec{\mathbf{J}}$ the total angular momentum operator, $\omega_{n \rightarrow m} = \epsilon_m-\epsilon_n$ the transition energy from the state $|n\rangle$ to the state $|m\rangle$ and $\mathcal{R}(\omega)$ the convolution of the Dirac $\delta-$function with an instrumental resolution function. The collective nature of the excitations being ignored at this stage, $S(|\vec{Q}|, \omega)$ depends on the scattering vector $\vec{Q}$ only through the magnetic form factor $f(|\vec{Q}|)$. 
Within the ground multiplet $^{5}$I$_8$ this can be computed in the dipole approximation as $f(|\vec{Q}|) = \langle j_0(|\vec{Q}|)\rangle + (g_L/g_J) \langle j_2(|\vec{Q}|)\rangle$ where $g_L = 3/4$ and $g_J = 5/4$ are the standard orbital and total gyromagnetic factors associated with the multiplet $^{5}$I$_8$. The radial integrals $\langle j_K(|\vec{Q}|)\rangle = \int_0^\infty r^2 R_{4f} j_K(r |\vec{Q}|) dr$, where $R_{4f}$ is the f$^{10}-$electron radial wavefunction and $j_K(r |\vec{Q}|)$ the Bessel function of order $K$, are tabulated in the 
International Tables of Crystallography \cite{ACL+06SM}. The fits of the CEF parameters were carried out by a reverse Monte Carlo method with Metropolis sampling and simulated annealing. The starting parameters were those reported for Ho$_2$Ti$_2$O$_7$ \cite{Ruminy16SM}. At high temperatures the decay of activated CEF excitations provided additional constraints. 

\subsubsection{Ir-Ho molecular exchange field}
Since no dispersion in the crystalline electric field excitations was detected within the accuracy of the experiments only upper bound for the exchange interactions can be estimated. A physical quantity that however revealed itself strongly sensitive to the Ir-Ho exchange interactions is the ordered magnetic moment $\vec{M}_{\rm Ho}$ of the Ho$^{3+}$ ions. This was extracted from the refinement, using an aiao order for the Ho sublattice, of the neutron diffraction experimental data and can be calculated at all temperatures as
\begin{equation}
\vec{M}_{\rm Ho}=g_J\mu_{\rm B} \Tr({\vec{\mathbf{J}} \exp({-\beta({\cal H}_{\rm CEF}+{\cal H}_{\rm Ho-Ir})}}))
\end{equation}
by writing the Hamiltonian $\mathcal{H}_{\rm Ho-Ir}$ of the  Ir-Ho exchange interactions in the Zeeman form
\begin{equation}
\mathcal{H}_{\rm Ho-Ir} = - \lambda \vec{M}_{\rm Ir}(T) \cdot g_J \mu_{\rm B}  \vec{\mathbf{J}}
\end{equation}
where $\lambda$ stands for the Ir-Ho molecular exchange field constant. The thermal variation of the Ir magnetic moment $\vec{M}_{\rm Ir}(T)$ is determined self-consistenly in its own Ir-Ir molecular exchange field, by assuming that it can be described as a spin $1/2$. The Ir-Ir molecular exchange field constant is determined by imposing that $\vec{M}_{\rm Ir}(T)$ cancels self-consistently at the Ir magnetic ordering temperature $T_{\rm MI} \approx$ 140~K. A similar processing of exchange interactions was previously implemented in Tb$_2$Ir$_2$O$_7$ \cite{Lefrancois15SM}.

\subsubsection{Refined parameters}
The best fits of the INS spectra at the different temperatures and for the different incoming neutron energies, on one hand, and of the thermal variation of the magnetic moments $\vec{M}_{\rm Ho}$, on the other hand, were obtained with CEF parameters $B_2^0 \approx 64$~meV, $B_4^0 \approx 242$~meV, $B_4^3 \approx 71$~meV, $B_6^0 \approx 89$~meV, $B_6^3 \approx -83$~meV, $B_6^6 \approx 90$~meV and a Ho-Ir molecular field $\lambda \vec{M}_{\rm Ir}(0) = 0.725(\pm0.050)$~T. The scattering function $S(|\vec{Q}|, \omega)$ in the neutron energy loss channel calculated at 1.5~K for incoming neutron energies $E_i = 48.4$~meV and $E_i = 101$~meV is displayed in the top panel of figure~\ref{fig:ExpSQwC}. 

The three main experimental signals are reproduced at the awaited energy transfer $\omega$ mimicking their decrease as a function of the scattering vector $|\vec{Q}|$. Their relative spectral weights differ to some extent from those of the experimental signals because of contributions from structural dynamics in the experiments. Additional weak signals are visible in the calculated spectra for incoming neutron energy $E_i = 101$~meV that accounts for CEF excitations at higher energy. The CEF leads to a ground doublet well separated from the excited levels that gives rise to a pseudo-Ising magnetic moment of magnitude $\approx$9.6~$\mu_{\rm B}$ oriented along the local three fold axis as in the Ho pyrochlore stannate and titanate \cite{Gardner10SM}. The parallel ($\|$) and perpendicular ($\bot$) Land\'e factors are calculated to g$_{\|} = 19.177$ and g$_{\bot} = 0.002$. 

The calculated thermal variation of the magnetic moment $\vec{M}_{\rm Ho}$ of the Ho$^{3+}$ ions induced by the Ho-Ir exchange field is displayed in the bottom panel of figure~\ref{fig:ExpSQwC} and compared to the experimental values extracted from the neutron diffraction data (see main article). They disagree below 20~K. No fitting of the whole experimental variation can be performed with only the Ho-Ir exchange field. As a matter of fact the temperature at which the measured and calculated moments depart from each other announces the onset of the magnetic correlations  associated with the exchange interaction between the Ho$^{3+}$ moments and the beginning of the fragmentation process.

%%%%%%%%%%%%%%%%%%%%%%%%%%%%%%%%%%%%%%%%
%\section{Dynamical properties}
%%%%%%%%%%%%%%%%%%%%%%%%%%%%%%%%%%%%%%%%

\section{Low temperature AC susceptibility}
\label{xac}

\begin{figure}[h!]
\includegraphics[width=8cm]{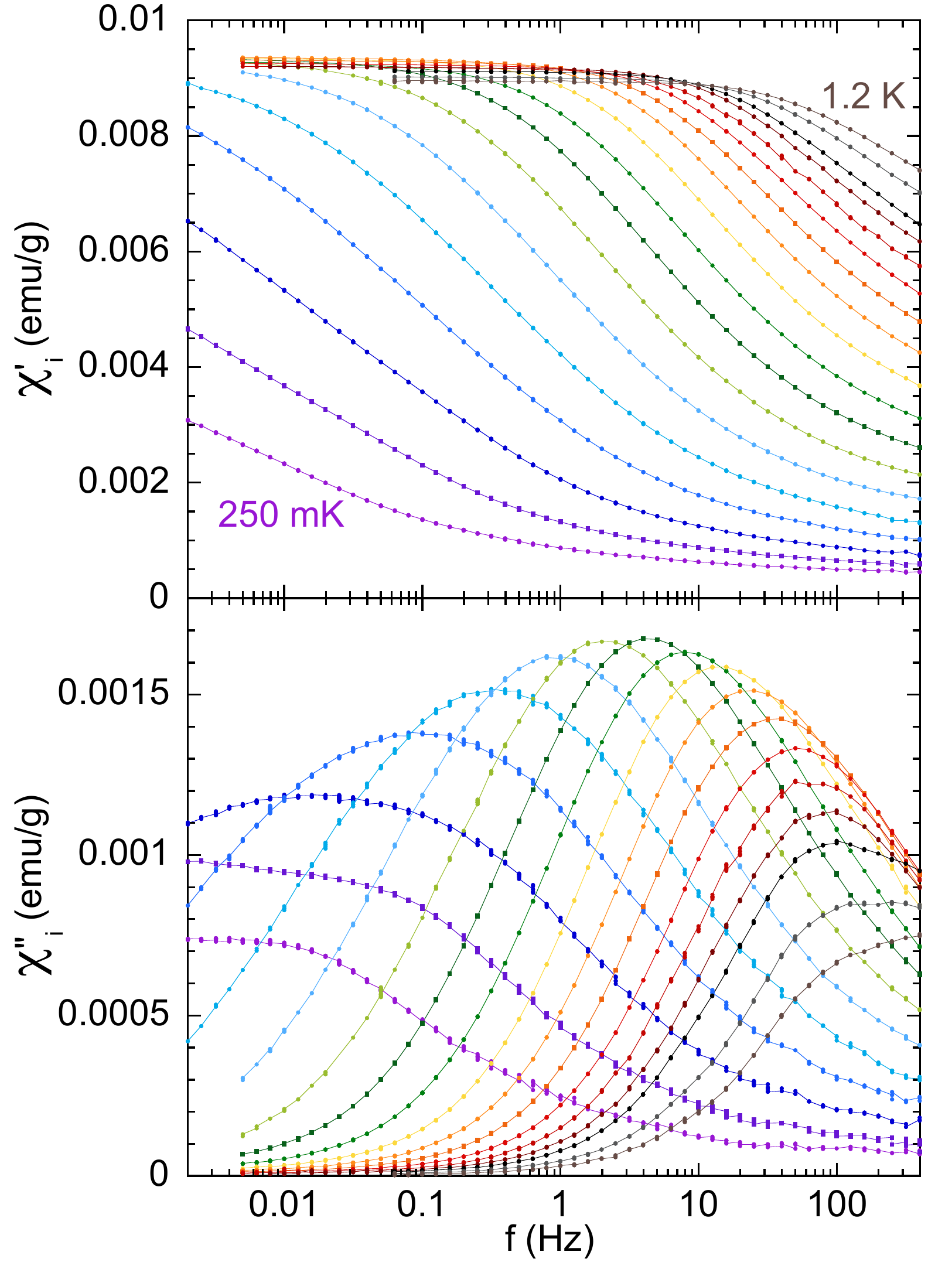}
\caption{\label{fig_Xvsf} Intrinsic ac susceptibility, corrected from demagnetizing effects, $\chi'_{\rm i}$ (top panel) and $\chi''_{\rm i}$ (bottom panel) as a function of frequency for temperatures between 250 mK and 1.2 K (by steps of 50 mK up to 1 K, and 100~mK above), measured with $H_{\rm ac}=1$~Oe. }
\end{figure}

AC susceptibility measurements in \HIO\ were performed at a fixed frequency $f$, as a function of temperature $T$ (as shown in the main text in figure 4), and at a fixed temperature, as a function of frequency (see figure \ref{fig_Xvsf}). The out-of-phase susceptibility $\chi''$ presents a maximum in both measurements, which is the signature of the presence of energy barriers that the system has to overcome when it is subjected to a magnetic field. Nevertheless, the curves are very broad, suggesting that the energy barrier is not unique, and that a distribution exists. In particular, the $\chi''$ vs $f$ curve does not display the lorentzian shape which is expected in presence of a single energy barrier. While at intermediate temperatures (typically $500-750$~mK), the curves can be fitted with a gaussian centered at a single frequency, at lower and higher temperatures, the shape is very asymmetric, indicating a distribution as well as several characteristic energies. For the lowest temperatures, due to the very long relaxation times, the system might have not reached its thermodynamic equilibrium, so that the energy distribution we have probed may be related to out-of-equilibrium processes. 

\begin{figure}[h!]
\includegraphics[width=8cm]{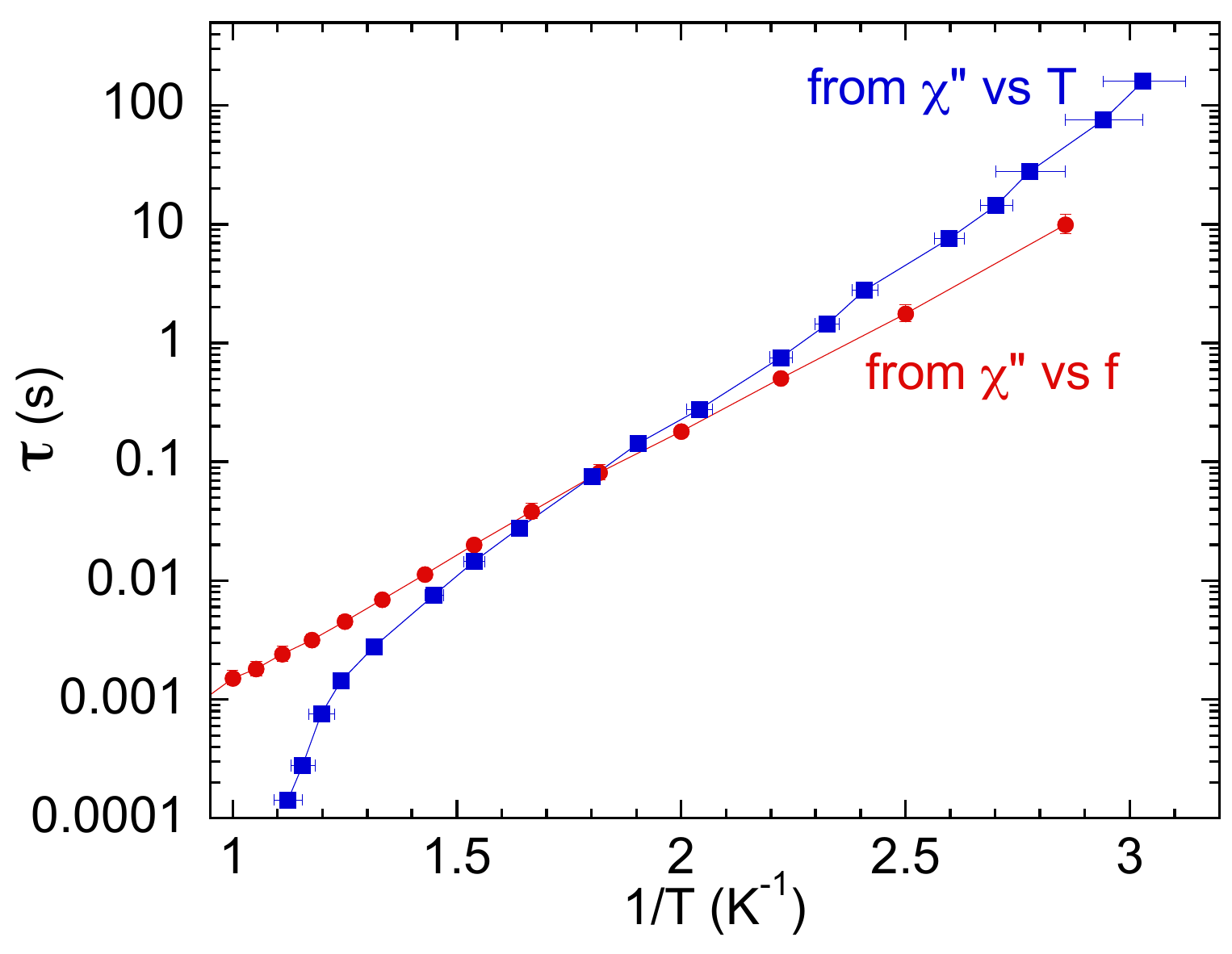}
\caption{\label{fig_tau} Relaxation time $\tau= 1 / 2\pi f$ vs $1/T$ determined from the temperature of the $\chi''$ maximum measured at constant frequency (blue squares) and from the frequency of the $\chi''$ maximum measured at constant temperature (red circles). }
\end{figure}

For these reasons, the maxima obtained from $\chi"$ vs $f$ and $\chi''$ vs $T$  measurements do not match, especially in the low and high temperature ranges (see figure \ref{fig_tau}), giving rise to different relaxation times $\tau$ (such discrepancy had already been reported in the spin ice Ho$_2$Ti$_2$O$_7$ \cite{Quilliam11SM}). Note that no relaxation time could be determined below 350 mK and above 1.1 K in $\chi''$ vs $f$ curves due to the broad shape of the curves. The ``true" relaxation time, defined in the context of Debye relaxation, is the one obtained from the $\chi''$ vs $f$ curves and it follows an Arrhenius law, as described in the main text. 

%%%%%%%%%%%%%%%%%%%%%%%%%%%%%%%%%%%%%%%%
%\section{Theoretical description of the dynamics}
%%%%%%%%%%%%%%%%%%%%%%%%%%%%%%%%%%%%%%%%

\section{Theoretical description of the dynamics}

\subsection{Charge model}

Below, we present the mapping of the spin model into a magnetic charge model, providing an alternative description of the local excitations (defects) and an insight into the origin of the energy barrier observed in ac susceptibility. 

\begin{figure}[t]
\includegraphics[width=8cm]{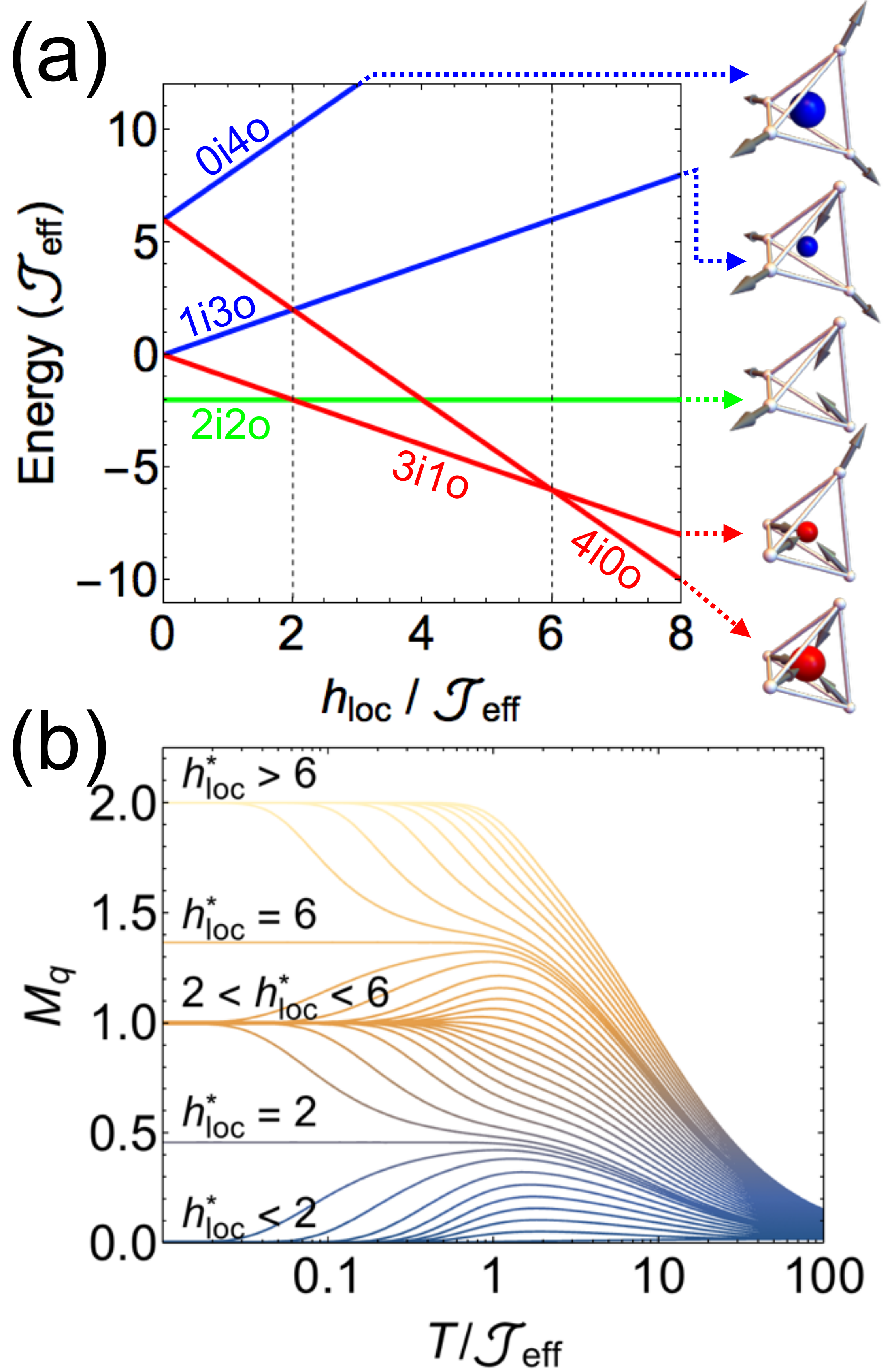}
\caption{\label{fig:diag} 
(a) Energies, in ${\cal J}_{\rm eff}$ units, of the different spin configurations of a single ``in" tetrahedron as a function of the local field $h^*_{\rm loc}=h_{\rm loc}/ {\cal J}_{\rm eff}$, showing the succession, when increasing the field, of the three different ground states 2i2o, 3i1o, and 4i0o. The red, blue and green colors indicate a positive, a negative and zero charge on a tetrahedron respectively. The small (resp. large) spheres represent charges $q=\pm 1$ (resp. $q=\pm 2$).
(b) Charge order parameter $M_q$ (defined in the main text) as a function of the temperature $T/{\cal J}_{\rm eff}$ for different local fields $h^*_{\rm loc}$.}
\end{figure}

Defining the charge $q_\alpha = \dfrac{\Delta_\alpha}{2} \sum_{i \in \alpha} \sigma_i$ on the tetrahedron $\alpha$, with $\Delta_\alpha = +1$, $-1$ for  ``in" and ``out" tetrahedra respectively, the following spin Hamiltonian
\begin{equation}
{\cal H} = {\cal J}_{\rm eff} \sum_{\langle i,j\rangle}{\sigma_i \sigma_j } - h_{\rm{loc}} \sum_{i} {\sigma_i},
\label{eqHs}
\end{equation} 
defined in the main text, can be written
\begin{equation}
{\cal H} = \sum_\alpha 2{\cal J}_{\rm eff} q_\alpha^2 - h_{\rm loc} \Delta_\alpha q_\alpha-2{\cal J}_{\rm eff}.
\label{eqHq}
\end{equation} 
From Equations (\ref{eqHs}-\ref{eqHq}), the different spin configurations of a given tetrahedron have the following energies (displayed on figure \ref{fig:diag}(a) as a function of the local field $h_{\rm loc}/ {\cal J}_{\rm eff}$): 

\begin{equation}
\begin{array}{lclcrcr}
E_{2i2o} & = & E_{q=0} & = & -2 {\cal J}_{\rm eff} & &         \\
E_{3i1o-1i3o} & = & E_{q=\pm 1} & = &       	   & \pm & \Delta_\alpha h_{\rm loc} \\
E_{4i0o-0i4o} & = & E_{q=\pm 2} & = & 6 {\cal J}_{\rm eff} & \pm & 2\Delta_\alpha h_{\rm loc}
\end{array}
\label{eqEnergies}
\end{equation}

Note that since each site is shared by two tetrahedra, so is the local field: the field term of equations (\ref{eqEnergies}) is then divided by a factor 2. 
This finally leads to the two phase boundaries, $h_{\rm loc}/{\cal J}_{\rm eff}=2$ ($E_{q=0} = E_{q=\pm 1}$) separating the 2i2o (spin ice) from the 3i1o--1i3o (fragmented) ground state, and  $h_{\rm loc}/ {\cal J}_{\rm eff}=6$ ($E_{q=\pm 1} = E_{q=\pm 2}$) separating the 3i1o--1i3o (fragmented) from the 4i0o--0i4o (ordered) ground state. 

These phase boundaries are confirmed by Monte Carlo simulations performed on extended pyrochlore lattices for different local fields $h_{\rm loc}/ {\cal J}_{\rm eff}$ and reduced temperatures $T/ {\cal J}_{\rm eff}$, as shown in figure \ref{fig:diag}(b) (as well as the figure 1(c) of the main text).

\subsection{Propagation of defects in the fragmented regime}

In spin ice materials, the diffusion of monopoles causes a slow relaxation of the magnetization which can be probed experimentally (see section \ref{xac} and main text). As a consequence, the relaxation timescale is proportional to the mobility of monopoles and inversely proportional to their density \cite{Jaubert09SM, Jaubert11SM, Castelnovo11SM}: the more defects in the system, the faster the relaxation.

In the fragmented regime ($2 < h_{\rm loc}/{\cal J}_{\rm eff} < 6$), the elementary excitations, although still fractional, are not canonical monopoles anymore: flipping a spin from the 3i1o--1i3o ground state now creates a pair of 2i2o (or 4i0o--0i4o) defects. We show in the following that the dynamics of these defects, propagating through different processes which depend on the value of $h_{\rm loc}$, is strongly unusual. 
\begin{figure*}[!btp]
\includegraphics[width=\textwidth]{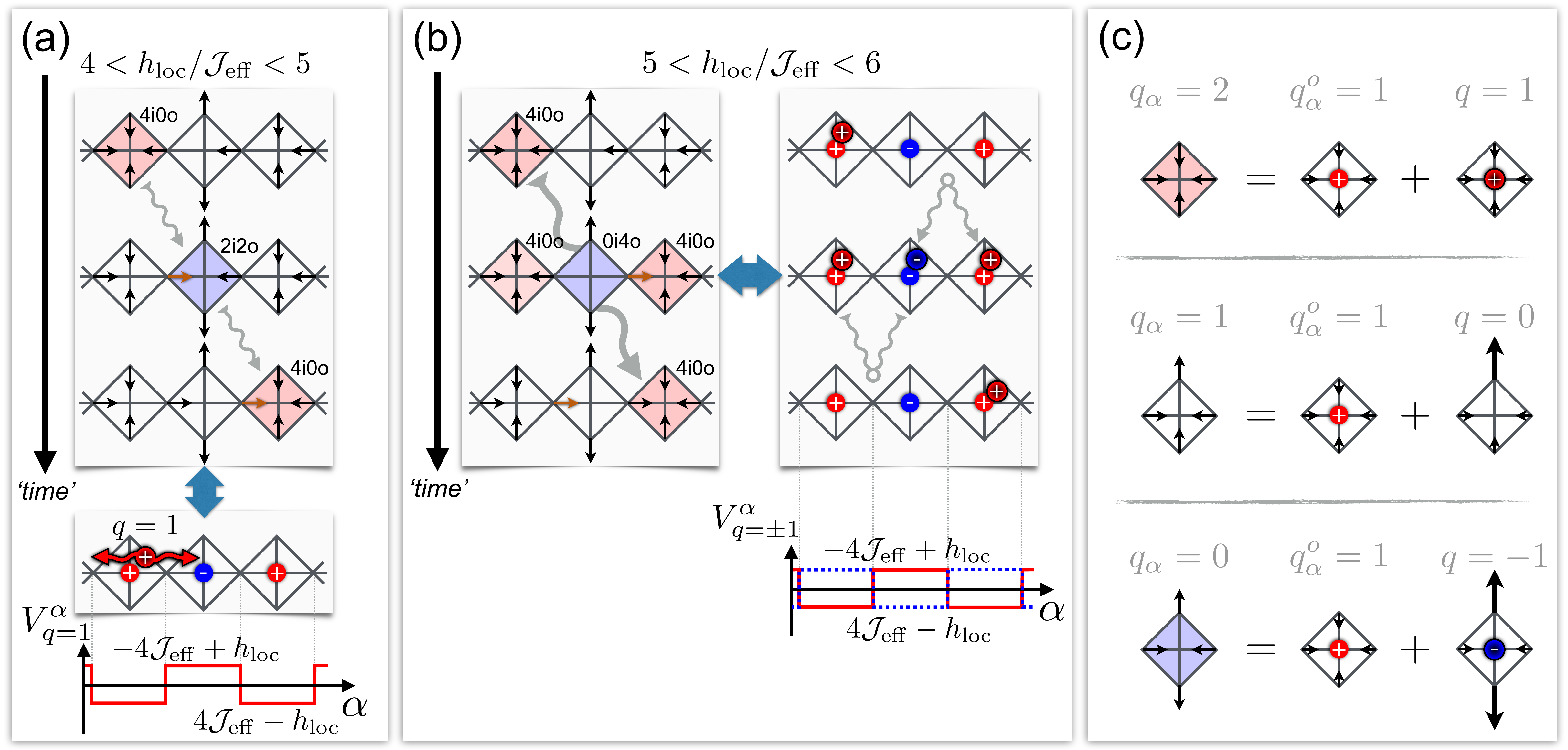}
\caption{\label{fig:defauts}
(a) Direct diffusion mechanism of a single defect carrying a positive charge $q=1$, dominant for $3<h_{\rm loc}/{\cal J}_{\rm eff}<5$, using spin {\it (top)} and charge {\it (bottom)} representations. The oscillating gray arrows show the motion of the defect at arbitrary times. The periodic potential at the bottom is shown for $4<h_{\rm loc}/{\cal J}_{\rm eff}<5$, and would be the opposite for $3<h_{\rm loc}/{\cal J}_{\rm eff}<4$. 
(b) Indirect diffusion mechanism of a single defect carrying a positive charge $q=1$, dominant for $5<h_{\rm loc}/{\cal J}_{\rm eff}<6$, using spin {\it (left)} and charge {\it (right)} representations. The gray arrows show the spontaneous creation ($t=1$) and annihilation ($t=2$) of two defects with opposite charges $q=\pm 1$. At bottom, the dashed blue line (resp. plain red line) represents the periodic potential of a negative (resp. positive) charge. The same mechanism occurs for $2<h_{\rm loc}/{\cal J}_{\rm eff}<3$ for 2i2o defects. 
(c) Mapping between the spin and the charge pictures for an ``in" tetrahedron.}
\end{figure*}

\subsubsection{Single excitation process}

We first focus on the most simple mechanism, dominant for $3<h_{\rm loc}/{\cal J}_{\rm eff}<5$, i.e. deep inside the fragmented regime, and which has been shortly discussed in the main text. It is schematized, on figure \ref{fig:defauts}(a), by the motion of a single defect along an unfolded string of tetrahedra. 
Its peculiarity comes from two (related) points: \\
(i) moving one defect from a tetrahedron to its neighbor by flipping one spin changes the nature of the defect: a 2i2o excitation transforms into a 4i0o (or 0i4o) excitations and vice versa, so that its propagation involves two alternating excitations.\\
(ii) the energies of the 2i2o and 4i0o--0i4o excitations, which depend on the local field $h_{\rm loc}$, are generally different (except for $h_{\rm loc}=4{\cal J}_{\rm eff}$) (see figure \ref{fig:diag}(a)), and so are their densities.

Interestingly, mapping this spin picture onto a charge description gives a better understanding of the dynamical properties of the fragmented regime, by characterizing the explicit role and behavior of each fragment: the defects are, in such a description, the elementary excitations (charges) of the Coulomb phase (divergence-free fragment), propagating in the periodic potential induced by the underlying static charge crystal (divergence-full fragment).

To show this, let us first consider that the charge $q_\alpha$ of the tetrahedron $\alpha$ is the combination of a static charge $q^\alpha_o = \Delta_\alpha$ (ordered charge of the charge crystal), and of a propagating charge $q$ (elementary excitation of the Coulomb phase). The $q$ value depends on the presence ($q=\pm 1$) or not ($q=0$) of a defect on the tetrahedron $\alpha$. 
The energy of such a defect is obtained by injecting $q_\alpha = q^\alpha_o + q$ in the Hamiltonian (\ref{eqHq}), and writes
\begin{equation}
E^\alpha_q = 2{\cal J}_{\mathrm{eff}} + V^\alpha_q,
\label{eqEq}
\end{equation} 
with the required energy to create a free charge $2{\cal J}_{\mathrm{eff}}$, and the potential energy $V^\alpha_q = 4{\cal J}_{\mathrm{eff}} q^\alpha_o q - h_{\mathrm{loc}} \Delta_\alpha q $. 
Noting that, in the fragmented regime,  $q^\alpha_o = \Delta_\alpha$, one can obtain the more simple expression $$V^\alpha_q = (4{\cal J}_{\mathrm{eff}} - h_{\mathrm{loc}}) \, q^\alpha_o \, q$$ which takes the form of an on-site interaction periodic potential induced by the underlying charge crystal, and is represented at the bottom of figure \ref{fig:defauts}(a) for a defect carrying a positive charge $q=+1$. For $h_{\rm loc}/{\cal J}_{\rm eff}>4$, charges of opposite (resp. similar) sign are repulsive (resp. attractive), while the opposite situation is realized for $h_{loc}/{\cal J}_{\rm eff}<4$.

\begin{figure}[!btp]
\includegraphics[width=8cm]{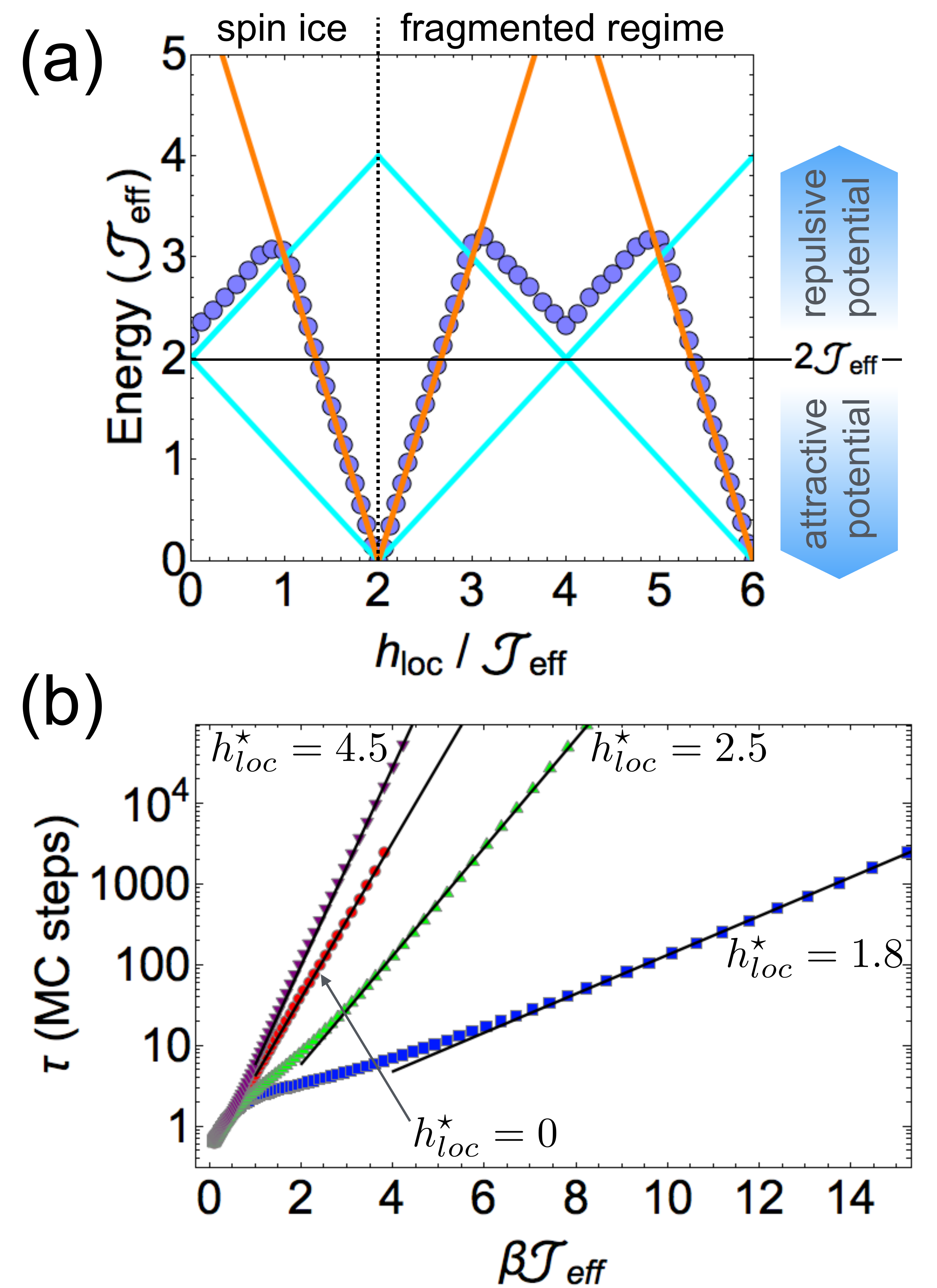}
\caption{\label{fig:MC}
(a) Characteristic energies associated to the energy barrier $\Delta E$ leading to the dynamical properties probed by ac susceptibility and setting the timescale $\tau = \tau_0 \exp(\beta \Delta E)$, as a function of the local field $h_{\rm loc}/{\cal J}_{\rm eff}$, obtained by Monte Carlo simulations for $N=8192$ spins (blue circles). The blue and orange lines represent single and multi-excitation processes respectively and their equations are given by the energies $E^\alpha_q$ of a single defect (equation \ref{eqEq}) and by $3E^\alpha_q$. As explained in the text, the single excitation process prevails for $3<h_{\rm loc}/{\cal J}_{\rm eff}<5$ and the multi-excitation process prevails close to the boundaries with the 2i2o ($2<h_{\rm loc}/{\cal J}_{\rm eff}<3$) and aiao ($5<h_{\rm loc}/{\cal J}_{\rm eff}<6$) phases.
(b) Characteristic relaxation timescale $\tau$ as a function of the inverse temperature $\beta$ for different values of the local field $h^*_{\rm loc}=h_{\rm loc}/ {\cal J}_{\rm eff}$. The symbols result from the fit of the correlation function $c(t)$ at each temperature, while the black lines represent the fit of $\tau$ vs $\beta$ (see text).}
\end{figure}

Coming back to a spin description of the dynamics, as shown in figure \ref{fig:defauts}(c), this means that the 2i2o defects (with a total charge $q^\alpha_o + q = 0$) have a lower energy than the 4i0o--0i4o ones  ($q^\alpha_o + q = \pm 2$) for $h_{\rm loc}/{\cal J}_{\rm eff}>4$, and the opposite situation for $h_{\rm loc}/{\cal J}_{\rm eff}<4$ (see figure \ref{fig:diag}(a)). The relaxation timescale $\tau = \tau_0 \exp(\beta \Delta E_1)$ in this single excitation regime is, finally, the time needed to overcome the energy barrier of the potential well $\Delta E_1 = h_{\rm loc} - 2{\cal J}_{\rm eff}$ (resp. $6{\cal J}_{\rm eff} - h_{\rm loc}$) for $h_{\rm loc}/{\cal J}_{\rm eff}>4$ (resp. $<4$) (see the cyan lines in figure \ref{fig:MC}(a)). 

\subsubsection{Multi-excitation process}

Approaching the critical fields $h_{\rm loc} = 2 {\cal J}_{\rm eff}$ and $6 {\cal J}_{\rm eff}$ in the fragmented state, the charge repulsion becomes such that another (more indirect but energetically favorable) diffusion process comes into play, accelerating the dynamics. As represented in figure \ref{fig:defauts} (b), this mechanism involves two successive spin flips. The first spin flip creates a pair of 2i2o (resp. 0i4o--4i0o) defects on the two neighboring tetrahedra of an already existing 2i2o (resp. 0i4o or 4i0o) defect. Then, the second spin flip simply annihilates one of these two defects with the old one, in such a way that this latter finally moved two tetrahedra away. This mechanism has the advantage to avoid the charge repulsion discussed earlier, by creating (close to an existing charge) two charges of opposite sign which both have an attractive potential with the charge crystal.

The energy cost for such a motion is, therefore, the energy of three charges trapped into their potential well $\Delta E_{2} = 18{\cal J}_{\rm eff} - 3h_{\rm loc}$ (resp. $3h_{\rm loc} - 6{\cal J}_{\rm eff}$) for $h_{\rm loc}/{\cal J}_{\rm eff}>4 $ (resp. $<4$). A direct comparison of $\Delta E_1$ and $\Delta E_2$ shows that the second mechanism, represented by the orange plain line in figure \ref{fig:MC}(a), is energetically favorable for $h_{\rm loc}/{\cal J}_{\rm eff}<3$ and $>5$.

\subsection{Numerical simulations}

This local field dependence of the charge dynamics has been confirmed by the numerical calculations of the autocorrelation function $c(t) = \frac{1}{N} \sum_i \langle \sigma_i(0) \sigma_i(t) \rangle$ using Monte Carlo simulations, with $t$ the Monte Carlo time, and $N = 8192$ the number of spins. In those simulations, the metropolis single spin flip algorithm, which mimics the local dynamics of the rare earth ions, has been used to evaluate $c(t)$, while it has been combined to a loop algorithm for thermalization. $c(t)$ has been fitted at each temperature and for different values of the local field within the range $2<h_{\rm loc}/{\cal J}_{\rm eff}<6$, assuming an exponential decrease of the autocorrelation with time $c(t) =\exp(-t/\tau)$. Then, the resulting characteristic timescale has been fitted as a function of the temperature, for $T/{\cal J}_{\rm eff}\lesssim 1$ and for each value of the local field, assuming an activation law $\tau = \tau_0 \exp (\beta \Delta E)$. This procedure, shown in figure \ref{fig:MC}(b) for several local fields, provides a numerical evaluation of the energy barrier $\Delta E$. The numerical simulations are in good agreement with the predictions discussed above, although the energy barriers obtained numerically for the single excitation process are slightly larger than expected.

Note also that the second mechanism is observed only at temperatures significantly lower than the energy difference between the two closest ground states (2i2o and 3i1o--1i3o around $h_{\rm loc}/{\cal J}_{\rm eff} =2$, and 3i1o--1i3o and 4i0o--0i4o around $h_{\rm loc}/{\cal J}_{\rm eff} =6$). At larger temperatures, these two spin configurations have similar Boltzmann weights and can coexist without a significant energy cost. Simple spin flips then allow the system to fluctuate between the two configurations.

Note finally that the case $h_{\rm loc} / {\cal J}_{\rm eff} < 2$, supporting a 2i2o Coulomb phase, has not been discussed above, although similar dynamics occur. The main difference with the fragmented regime is the absence of charge order, such that the periodic potential is here only created by the staggered local field $V^\alpha_q = - h_{\mathrm{loc}} \Delta_\alpha q $ which plays a role comparable to the charge crystal in the fragmented regime.

\begin{figure*}[!t]
\includegraphics[width=\textwidth]{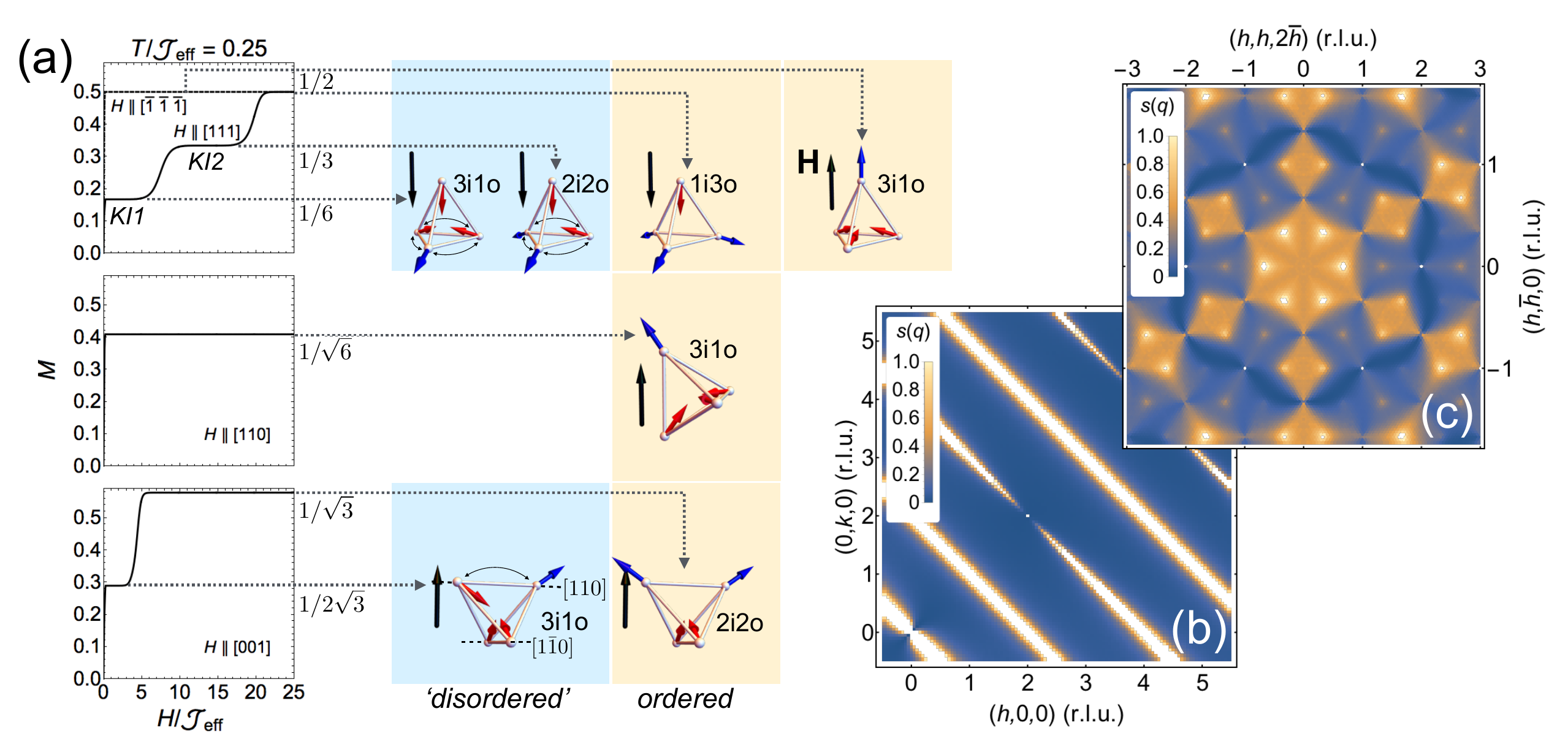}
\caption{\label{fig:mhs} (a) Magnetization $M$ per spin (assuming $S=1$), obtained from Monte Carlo simulations for $N=8192$ spins, as a function of the global field $H$ applied along the $[111]$ and $[\bar{1}\bar{1}\bar{1}]$ directions (plain and dashed curve of the top panel), along the $[110]$ direction (middle panel), and along the $[001]$ direction (bottom panel). The value of the magnetization on each plateau is indicated on the right of the plots, together with a scheme of the stabilized spin configurations considering an ``in" tetrahedron. Note that inverting the field from $[111]$ to  $[\bar{1}\bar{1}\bar{1}]$ is equivalent to going from one aiao iridium 180$^{\circ}$ domain to the other one. KI1 and KI2 label the successive fragmented kagome ice phases stabilized for $H \parallel [\bar{1}\bar{1}\bar{1}]$. Their scattering function, exhibiting Bragg peaks and pinch point pattern, is represented in (c). The scattering function of the partially disordered state stabilized for $H \parallel [001]$, shown in (b), displays rods of diffuse scattering.}
\end{figure*}

%%%%%%%%%%%%%%%%%%%%%%%%%%%%%%%%%%%%%%%%
%\section{Magnetization in the presence of an external field}
%%%%%%%%%%%%%%%%%%%%%%%%%%%%%%%%%%%%%%%%

\section{Magnetization in the presence of an external field}

As noted in the main text, the measured magnetization of \HIO\ in an external magnetic field $\mathbf{H}$ presents an unconventional behavior. Due to the powder average, these data are difficult to analyze quantitatively. However, we anticipated that rich plateau physics should underlie our observations. We have thus performed Monte-Carlo calculations for $h_{\rm loc}/{\cal J}_{\rm eff} = 4.5$, the value we have estimated for \HIO, to get a first insight into the involved physics. These calculations indeed disclose the stabilization of field-induced exotic phases. 

When the field is applied along the [111] direction, for which the pyrochlore lattice consists of alternating  kagome and triangular planes, we get two successive magnetization plateaus KI1 and KI2 (see Figure \ref{fig:mhs}). They correspond to kagome ice phases, i.e. fragmented phases in the kagome planes supporting a ``2i1o--1i2o" charge crystal together with fluctuating disordered spins, while the spins of the triangular planes are aligned along the magnetic field. The corresponding scattering function results in the coexistence of a diffuse signal presenting pinch points and of Bragg peaks associated with both the charge order in the kagome planes with a propagation vector $\mathbf{k} = (2/3,2/3,0)$ and the polarized spins of the triangular lattice with $\mathbf{k} = 0$ \cite{Bartlett14SM} (see figure \ref{fig:mhs}(c)).  

Although KI1 and KI2 stabilize the same charge order in the kagome planes, they support distinct three dimensional charge orders, with different rules on each tetrahedron and different magnetization values. Those different spin configurations reflect the competition between the staggered field $h_{\rm loc}$ and the increasing external field $H$. When $H$ is small (KI1), the field-induced state on the first plateau still corresponds to a charge crystal on the pyrochlore lattice, made of 3i1o and 1i3o configurations: the spins in the triangular planes are polarized by the applied field, and the kagome ice forms in the kagome planes with one spin having a positive projection along the field and the two others a negative one, so that the global rules 3i1o (or 1i3o) on the tetrahedra are preserved (see upper panels of Figure \ref{fig:mhs}(a) for an ``in" tetrahedron). This results in a magnetization plateau at $M  =  \dfrac{1}{N} \mathbf{H} \cdot \left( \sum_i \mathbf{S}_i \right) = \frac{1}{4} \left(  1 + \frac{1}{3} - \frac{1}{3} - \frac{1}{3} \right) = \frac{1}{6}$ with $N$ the number of spins and assuming $S=1$. 

When $H$ increases (KI2), the Zeeman energy is such that it becomes favorable, in each triangle of the kagome plane, to flip a spin so that its projection along the field goes from negative to positive. Each tetrahedron then stabilizes a 2i2o spin configuration, still inducing a kagome ice state, but leading to a global jump in the magnetization curve towards a plateau at $M  = \frac{1}{4} \left(  1 + \frac{1}{3} + \frac{1}{3} - \frac{1}{3} \right) = \frac{1}{3}$. Note that the phase KI2 is the same as in a conventional spin ice under a [111] applied field \cite{Matsuhira02SM, Tabata06SM}. KI1 and KI2, having exactly the same correlations in the kagome planes, present the same scattering function as well (identical diffuse scattering and Bragg peaks of the charge crystal). Actually, they only differ by the intensity of their Bragg peaks at Brillouin Zone centers.

At larger field, the system further polarizes in the $\mathbf{H}$ direction leading to an ordered 3i1o--1i3o state with $M = \frac{1}{4} \left(  1 + \frac{1}{3} + \frac{1}{3} + \frac{1}{3} \right) = \frac{1}{2}$. This plateau succession only concerns one out of the two possible 180$^{\circ}$ spin domains, themselves selected by the underlying 180$^{\circ}$ magnetic domains of the aiao iridium magnetic order. The second domain $-$ which can be obtained by reversing all iridium spins and thus all the local fields $-$ is immediately polarized in a 1i3o--3i1o ordered phase when a [111] external field is applied, as shown by the dashed curve in the upper panel of figure \ref{fig:mhs}(a). Note that inverting the local fields or the global applied field produces the same effect, so that this dashed curve has been labelled $H\parallel [\bar{1}\bar{1}\bar{1}]$.

For a field applied along the [001] direction, a magnetization plateau is also observed at half the saturation $M  = \frac{1}{4\sqrt{3}} \left(  1 +1 + 1 -1 \right) = \frac{1}{2\sqrt{3}}$, associated with long range 1D correlations along the spin chains perpendicular to the field. These spin chains can be separated into two sets: one, fully ordered, has all its spins preferentially oriented in the field direction (chains along $[1\bar{1}0]$ in figure \ref{fig:mhs}(a)); the second set, characterized by alternating in and out spins along the $[110]$ chains, remains disordered from chain to chain. Therefore, the resulting scattering function in the $(h, k, 0)$ scattering planes shows rods of diffuse scattering along $[1\bar{1}0]$ together with Bragg peaks at Brillouin Zone centers. Note that this state is similar to the spin ice case with a field applied along the $[110]$ direction \cite{Hiroi03SM, Fennell05SM}.

Finally, for a field applied along the [110] direction, the magnetization saturates very fast toward a 3i1o--1i3o ordered state with $M = \frac{1}{\sqrt{6}}$. 
 
%%%%%%%%%%%%%%%%%%%%%%%%%%%%%%%%%%%%%%%%
%Bibliography
%%%%%%%%%%%%%%%%%%%%%%%%%%%%%%%%%%%%%%%%

\end{document}